\global\def\draftcontrol{0}

   \def\versionno{ n2bh  curved horizon}

\catcode`\@=11

\expandafter\ifx\csname draftcontrol\endcsname\relax\global\def\draftcontrol{0}
\fi

{\count255=\time\divide\count255 by 60
\xdef\hourmin{\number\count255}
\multiply\count255 by-60\advance\count255 by\time
\xdef\hourmin{\hourmin:\ifnum\count255<10 0\fi\the\count255}}
\def\draftdate{\number\month/\number\day/\number\year\ \ \ \hourmin }

\newcommand\makepapertitle{\par
  \begingroup
    \renewcommand\thefootnote{\@fnsymbol\c@footnote}%
    \def\@makefnmark{\rlap{\@textsuperscript{\normalfont\@thefnmark}}}%
    \long\def\@makefntext##1{\parindent 1em\noindent
            \hb@xt@1.8em{%
                \hss\@textsuperscript{\normalfont\@thefnmark}}##1}%
     \newpage
     \global\@topnum\z@   
     \@makepapertitle
     \thispagestyle{empty}\@thanks
  \endgroup
  \setcounter{footnote}{0}%
  \global\let\thanks\relax
  \global\let\makepapertitle\relax
  \global\let\@makepapertitle\relax
  \global\let\@thanks\@empty
  \global\let\@author\@empty
  \global\let\@date\@empty
  \global\let\@title\@empty
  \global\let\title\relax
  \global\let\author\relax
  \global\let\date\relax
  \global\let\and\relax
  \def\version{\let\version\@version\@gobble}
}
\def\@makepapertitle{%
  \newpage
   \ifnum\draftcontrol=1 {}
   \version\versionno
   \vskip 3em%
   \else
   \hfill\hbox to 3cm {\parbox{4cm}{\@pubnum}\hss}%
   \vskip 3em%
   \fi
   \begin{center}%
   \let \footnote \thanks
     {\LARGE {\@title}}%
     \vskip 1.5em%
     {\normalsize
       \lineskip .5em%
       \begin{tabular}[t]{c}%
         \@author
       \end{tabular}\par}%
     \vskip 1.5em%
     {\@bstract}%
     \end{center}%
     \vskip 1.5em
     \@date%
   \par
}

\gdef\@pubnum{}
\def\pubnum#1{%
  \gdef\@pubnum{#1}}

\gdef\@bstract{}
\def\Abstract#1{%
  \gdef\@bstract{%
   \parbox{\textwidth-0pc}{%
   \centerline{\bf Abstract}\penalty1000%
\kern.2cm%
\noindent
\renewcommand\baselinestretch{1.0}%
{#1}}}
}

\def\ps@paper{\let\@mkboth\@gobbletwo%
     \ifnum\draftcontrol=1
    \def\@oddfoot{\hbox to \textwidth{\tiny \versionno \hfil\tiny\draftdate}%
    \hskip -\textwidth \hbox to \textwidth{\hfil\rm\thepage\hfil}}%
     \else\def\@oddfoot{\hbox to \textwidth{\hfil\rm\thepage\hfil}}
     \fi
     \let\@evenfoot\@oddfoot
}

\def\body{\clearpage
          \pagestyle{paper}
    }

\def\@version#1{\ifnum\draftcontrol=1
\typeout{}\typeout{#1}\typeout{}
\vskip3mm\centerline{\hbox{\fbox{\normalsize{\tt DRAFT -- #1 -- }
                   {\draftdate}}}}\vskip3mm
\fi}
\let\version\@version
\long\def\eqlabel#1{\ifnum\draftcontrol=1
                    \tag@false  
                    \tag*{(\theequation) \hbox to -0.2cm{\hspace{0cm}\small{#1}\hss}}
                    \refstepcounter{equation}
                    \edef\@currentlabel{\theequation}
                    \ltx@label{#1}          
                    \else
                    \label{#1}
                    \fi
                    }
\let\st@bibitem\@bibitem
\let\st@lbibitem\@lbibitem
\ifnum\draftcontrol=1
  \def\@bibitem#1{%
    \st@bibitem{#1}\a@@label{#1}\ignorespaces}
  \def\@lbibitem[#1]#2{%
    \st@lbibitem[#1]{#2}\a@@label{#2}\ignorespaces}
  \def\a@@label#1{%
    \gdef\a@lab{\smash{\normalfont\small#1}}
    \ifvmode
      \if@inlabel
        \global\setbox\@labels\hbox{%
          \llap{\a@lab\let\a@lab\relax
                \kern\@totalleftmargin\kern\marginparsep}%
          \box\@labels}%
      \fi
    \fi}
\fi

\documentclass[12pt,letterpaper]{article}

\usepackage{amsmath,amssymb,array,calc,epsfig,rotating,bm,xcolor}
\usepackage[sort]{cite}
\usepackage{graphicx,esint,float}
\usepackage{psfrag,verbatim}
\usepackage[makeroom]{cancel}
\usepackage{xcolor,url}
\usepackage{hyperref}

\ifnum\draftcontrol=1
\fi

\renewcommand\baselinestretch{1.25}
\renewcommand\section{\@startsection {section}{1}{\z@}%
                                   {-3.5ex \@plus -1ex \@minus -.2ex}%
                                   {2.3ex \@plus.2ex}%
                                   {\normalfont\large\bfseries}}
\renewcommand\subsection{\@startsection{subsection}{2}{\z@}%
                                   {-3.25ex\@plus -1ex \@minus -.2ex}%
                                   {1.5ex \@plus .2ex}%
                                   {\normalfont\normalsize\bfseries}}
\renewcommand\subsubsection{\@startsection{subsubsection}{3}{\z@}%
                                   {-3.25ex\@plus -1ex \@minus -.2ex}%
                                   {1.5ex \@plus .2ex}%
                                   {\normalfont\normalsize\it}}
\renewcommand\paragraph{\@startsection{paragraph}{4}{\z@}%
                                   {-3.25ex\@plus -1ex \@minus -.2ex}%
                                   {1.5ex \@plus .2ex}%
                                   {\normalfont\normalsize\bf}}

\numberwithin{equation}{section}

\def\revise#1       {\raisebox{-0em}{\rule{3pt}{1em}}%
                     \marginpar{\raisebox{.5em}{\vrule width3pt\
                     \vrule width0pt height 0pt depth0.5em
                     \hbox to 0cm{\hspace{0cm}{%
                     \parbox[t]{4em}{\raggedright\footnotesize{#1}}}\hss}}}}

\newcommand\nxt[1]  {\\\fnxt#1}
\newcommand{\ie}{{\it i.e.,}\ }
\newcommand{\eg}{{\it e.g.,}\ }
\newcommand{\mt}[1]{\textrm{\tiny #1}}

\def\calc         {{\cal C}}
\def\cald         {{\cal D}}
\def\cale         {{\cal E}}
\def\calf         {{\cal F}}

\def\call         {{\cal L}}
\def\calm         {{\cal M}}
\def\caln         {{\cal N}}
\def\calo         {{\cal O}}

\def\reals        {{\mathbb R}}
\def\zet          {{\mathbb Z}}

\def\del          {\partial}

\def\tr           {\mathop{\rm Tr}}
\def\Re           {{\rm Re\hskip0.1em}}
\def\Im           {{\rm Im\hskip0.1em}}

\def\sqr#1#2{{\vcenter{\vbox{\hrule height.#2pt
 \hbox{\vrule width.#2pt height#1pt \kern#1pt
 \vrule width.#2pt}\hrule height.#2pt}}}}
\def\square{%
  \mathop{\mathchoice{\sqr{12}{15}}{\sqr{9}{12}}{\sqr{6.3}{9}}{\sqr{4.5}{9}}}}

\newcommand{\kk}{\mathfrak{q}}
\newcommand{\ww}{\mathfrak{w}}

\def\kt{\tilde{k}}

\def\aa1{\phi}
\def\cc1{\psi}

\def\tk{\tilde{k}}

\def\f0{\text{\boldmath$\varphi$}}
\def\h2{\mathfrak{h}}

\catcode`\@=12

\begin{document}


\title{\bf Stabilization of the extended horizons}

\date{May 27, 2021}

\author{
Alex Buchel\\[0.4cm]
\it $ $Department of Applied Mathematics\\
\it $ $Department of Physics and Astronomy\\ 
\it University of Western Ontario\\
\it London, Ontario N6A 5B7, Canada\\
\it $ $Perimeter Institute for Theoretical Physics\\
\it Waterloo, Ontario N2J 2W9, Canada
}

\Abstract{We extend Kodama-Ishibashi and Jansen-Rostworowski-Rutkowski
master field framework to study quasinormal modes of black branes and
black holes of Einstein gravity in $D=5$ space-time dimensions
with multiple scalars and an arbitrary bulk potential.
As an application, we consider a thermodynamically unstable state of
mass deformed $\caln=4$ supersymmetric Yang-Mills theory at strong coupling.
The  corresponding gravitational dual is $\caln=2^*$ black brane
with a dynamical instability of its translationary invariant horizon.
We study the dependence of the quasinormal mode spectra as the boundary
gauge theory is compactified on $S^3$ --- the black brane is turned into
the black hole.  
}

\makepapertitle

\body

\version\versionno
\tableofcontents

\section{Introduction}\label{intro}

One of the thermodynamic inequalities of a thermal state at
equilibrium,
also known as a condition for a  thermodynamic stability, is
\cite{Landau:1980mil},
\begin{equation}
  \frac{c_v}{s}\equiv \frac Ts\ \left(\frac{\del s}{\del T}\right)_v \ >\ 0\,,
\eqlabel{therstab}
\end{equation}
where $T$ is a temperature density, $s$ is an entropy density
and $c_v$ is a constant volume specific heat density. 
In gauge theory/string theory correspondence
\cite{Maldacena:1997re,Aharony:1999ti}, the  properties of thermal states of
strongly coupled gauge theories are encoded in thermodynamics of the
corresponding planar black hole horizon geometries. The thermodynamic
stability is equally applicable to black branes: here, the entropy density
is the Bekenstein entropy density of a regular Schwarzschild planar horizon,
and the temperature is its Hawking temperature.  Perturbed black branes
relax to thermal equilibrium via quasinormal modes (QNMs)
\cite{Berti:2009kk}. In some cases the spectra of QNMs contains
unstable states, \ie the QNMs with\footnote{We define
reduced frequency and momenta as $\ww\equiv w/(2\pi T)$ and $\kk=k/(2\pi T)$.}
\begin{equation}
\Im[\ww] > 0\,.
\eqlabel{dynsta}
\end{equation}
When excited, these modes grown exponentially (in a linearized approximation),
rather then decaying. In the dual boundary language, their presence signals
a dynamical instability of the strongly coupled plasma  to
density-pressure fluctuations breaking the translational invariance
of the thermal equilibrium state.

In \cite{Gubser:2000ec,Gubser:2000mm}
the authors formulated the correlated stability conjecture (CSC):
 \begin{center}
    {\it A black brane thermodynamic (in)stability correlates with its
      dynamical (in)stability}
    \end{center}
The conjecture is applicable only to horizons with translational
invariance, \eg simple higher-dimensional Schwarzschild black holes
in Einstein gravity with a cosmological constant are
stable \cite{Ishibashi:2003ap}. If true, the CSC would provide a
way to bypath technically complicated QNM analysis and reestablish
the dynamical (in)stability of a horizon. 

In one direction, the CSC is trivial \cite{Buchel:2005nt}:
thermodynamic instability of an extended horizon implies its dynamical
instability. Indeed, a basic thermodynamic relation between
the speed of the sound waves in (uncharged) plasma $c_s$ and its specific heat
\begin{equation}
c_s^2=\frac{s}{c_v}\,,
\eqlabel{cscv}
\end{equation}
implies that when $c_v<0$ the speed of the sound waves is purely imaginary.
As a result, the sound mode, \ie the scalar channel black
brane QNM,
\begin{equation}
\ww(\kk)=\pm c_s\ \kk  -2\pi i \ \frac{\eta}{s}\left(\frac 23
+\frac{\zeta}{2\eta}\right)\ \kk^2+\calo(\kk^3)\,,
\eqlabel{sound}
\end{equation}
is unstable for some sign and for small enough $\kk$.
In \eqref{sound}, $\eta$ and $\zeta$ are the shear and the bulk viscosities
of the plasma.  The presence of such dynamical instability in
thermodynamically unstable horizons was explicitly verified in
\cite{Buchel:2005nt,Buchel:2008uu,Buchel:2010gd,Buchel:2010wk}.

In the other direction the conjecture is simply wrong\footnote{See
\cite{Buchel:2010wk,Buchel:2011ra} for an earlier work.}.  A
sharp recent counterexample is the holographic conformal
order \cite{Buchel:2020thm,Buchel:2020xdk,Buchel:2020jfs}.
It is straightforward to 
construct holographic models in asymptotically
    AdS$ _{d+2}$ with, say $\zet_2$ global symmetry, such that
\begin{equation}
    \frac{\calf}{T^{d+1}}=-\ \calc\ \times\
    \begin{cases}
      1,\qquad \langle\calo\rangle=0\Longrightarrow \zet_2\
      {\rm is\ unbroken};\\
      \kappa,\qquad \langle\calo\rangle\ne 0\Longrightarrow \zet_2\
      {\rm is\ spontanuously\ broken}\,,
      \end{cases}
      \eqlabel{conforder}
\end{equation}
    where $\calf$ is the free energy density,
    $\calc$ is a positive constant proportional to the central charge
    of the theory, and $0<\kappa<1$ is also a constant.
    The thermodynamics and the hydrodynamics of both the symmetric
    and the symmetry broken phases
    are identical, \ie that of the CFT$ _{d+1}$. In particular, see
    \eqref{therstab},
    \begin{equation}
    \frac{c_v}{s}=d\ >\ 0\,,
    \eqlabel{cvsorder}
    \end{equation}
\ie both phases are thermodynamically stable. On the other hand,
there is \cite{Buchel:2020jfs} a (non-hydrodynamic) branch of the scalar sector
    QNMs that renders the symmetry broken phase {\it perturbatively unstable},
    \ie there is a QNM of the type \eqref{dynsta} at $\kk=0$.

The motivation of this study is to 'fix' the instabilities of the extended
horizons, by compactifying the Euclidean space $\reals^d$
of a dual holographic CFT$ _{d+1}$
(QFT$ _{d+1}$ more generally) on $d$-dimensional round sphere $S^d$:
\begin{equation}
 \reals^d\ \longrightarrow\ L^2\ S^d\,,\qquad K\equiv \frac{1}{L^2}\,.
\eqlabel{compactd}
\end{equation}
Specific potential applications are
\begin{itemize}
\item  Curing the instabilities in the hydrodynamic sector,
\ie in models with $c_s^2 <0$. An example:
Klebanov-Strassler black branes \cite{Buchel:2018bzp}.
  \item Curing the instabilities in the non-hydrodynamic sector, \ie in models
  of the holographic conformal order \cite{Buchel:2020jfs}. 
\end{itemize}

The rest of the paper is organized as follows.
In appendix \ref{eomshol} we extend the QNM analysis of
\cite{Kodama:2003jz,Jansen:2019wag} to black branes/holes in theories
of Einstein gravity in $D=5$ space-time dimensions
with multiple scalars and an arbitrary bulk potential:
\begin{equation}
   S_5=\int_{\calm_5}d^{5}\xi\ \sqrt{-g} \biggl[R-\sum_{j=1}^p \eta_j \left(\del\phi_j\right)^2-V\left(\{\phi_j\}\right)\biggr]\,.
\eqlabel{potdub}
\end{equation}
We then apply the general formalism to $\caln=2^*$ holographic model
\cite{Pilch:2000fu,Buchel:2000cn,Evans:2000ct}, summarizing the
results in section \ref{summary}. Technical details of the
computations are discussed in section \ref{tec}.
We conclude in section \ref{conclude}. Summary
of the numerical tests performed is presented in appendix \ref{err}.

Thermodynamically unstable
phase of $\caln=2^*$ plasma was identified in \cite{Buchel:2007vy}.
The hydrodynamics of this phase, and its dynamical instability, was
discussed in \cite{Buchel:2007mf,Buchel:2008uu,Buchel:2009hv}. Detailed analysis
of the $S^3$-compactified $\caln=2^*$ unstable thermodynamics is
new\footnote{Some early work appeared in \cite{Buchel:2003qm,Buchel:2015lla}.}.
Computation of the QNM spectra of $\caln=2^*$ black holes is new.

\section{Summary}\label{summary}
In this section we present
results\footnote{Also available as a recent talk
\cite{talk}.} of the case study of the extended unstable horizons
in $\caln=2^*$ holographic model. We focus on a particular top-down holographic model,
and it is interesting to explore in the future how generic the reported results are. 
It is a model of a {\it real} string holography
\cite{Buchel:2000cn,Buchel:2013id,Bobev:2013cja,Bobev:2018hbq} --- we feel it is important
to emphasize this as sometimes subtle effects in toy holographic models do not occur
in string theory \cite{Buchel:2020xdk,inprep1}.

We focus on the thermodynamically unstable phase of the $\caln=2^*$ theory, where a single
mass parameter $m$  is introduced to the bosonic components of the $\caln=2$ hypermultiplet
of $\caln=4$ supersymmetric $SU(N)$ Yang-Mill theory. We work in the
planar limit of the gauge theory, and at large 't Hooft coupling --- in this case the full
string theory/gauge theory duality is reduced to a supergravity approximation
\cite{Maldacena:1997re}.
The full ten dimensional Type IIB supergravity can be consistently truncated on $S^5$, producing
the Pilch-Warner (PW) effective action \cite{Pilch:2000fu}. The latter effective action is in the
general class of models covered in appendix \ref{eomshol}. 

The general plan is as follows.
\nxt \underline{\bf(A)} We begin with the $\caln=2^*$ black brane thermodynamics and
hydrodynamics.
\begin{itemize}
\item We identify a thermodynamically unstable state of 
$\caln=2^*$ model at $K=0$ --- this is a purple
dot\footnote{There is nothing special with our selection of the
thermodynamically  unstable state ---
any state on the red dashed curves would do. The results are
qualitatively the same.} in fig.~\ref{macro},
and the left panel of fig.~\ref{micro}.
\item Next, we compute the hydrodynamic properties of this state ---
fig.~\ref{csbv}. We establish that the speed of the sound waves $c_s^2<0$,
and compute the bulk viscosity $\zeta$ (the purple dots,
both panels). We use the hydrodynamic limit
\eqref{sound} as an independent
check on the computation of the QNM spectra
in the framework of appendix \ref{eomshol} --- the thin red dashed
curves in fig.~\ref{hydroK0}.
\item The sound channel QNM of the selected state (the purple dot)
is computed in fig.~\ref{hydroK0}.
We establish the expected \cite{Buchel:2005nt} dynamical instability.
\end{itemize}
$\ $
\nxt \underline{\bf(B)} We continue with the non-hydrodynamic QNMs of the $\caln=2^*$ black
brane. 
\begin{itemize}
\item We compute the spectrum of the non-hydrodynamic modes in the
helicity $h=0$ sector at $\kk=0$, initially at $m^2=0$, \ie in pure AdS$ _5$-Schwarzschild black brane,
see fig.~\ref{adsqnm}. This is one of the two branches of the QNMs
in this helicity sector --- the other branch contains the hydrodynamic
(sound) mode with $\ww(\kk=0)=0$, detailed in fig.~\ref{hydroK0}.
The two branches decouple at $\kk=0$. We highlight the two lowest
QNMs: the magenta and the pink one, which we follow, increasing
$\frac{m^2}{T^2}$, to the select unstable thermal state of interest
at \eqref{tm}.
\item The dependence of the magenta (the lowest in  fig.~\ref{adsqnm})
QNMs at $\kk=0$ as $\frac{m^2}{T^2}$ varies as
\begin{equation}
0\leq\frac{m^2}{T^2}\leq \frac{m^2}{T_{p}^2}\,,
\eqlabel{flowqnm}
\end{equation}
is shown in fig.~\ref{f0mode1K0}.
Fig.~\ref{f0mode2K0} tracks over the same mass range the pink QNMs in
fig.~\ref{adsqnm}. Both modes remain stable, \ie have $\Im[\ww]<0$,
over the full mass range \eqref{flowqnm}.
\end{itemize}
$\ $
\nxt \underline{\bf(C)} We proceed introducing the curvature to the black brane
horizon of the select unstable state. We omit the description of the background,
and present only the results for the QNMs at temperature \eqref{tm} as
$\frac{K}{m^2}$ varies.  A sample background thermodynamics
at $\frac{K}{m^2}=1$ is shown in fig.~\ref{macro1} and the right panel of
fig.~\ref{micro}.

In the absence of curvature, the dispersion
of the low energy quasinormal modes is characterized by  few
transport coefficients (the speed of the sound waves $c_s$, the shear
$\eta$ and the bulk $\zeta$ viscosities) --- see \eqref{sound}
and fig.~\ref{hydroK0} ---  valid, provided 
\begin{equation}
\kk\equiv \frac{k}{2\pi T}\ll 1\qquad {\rm and}\qquad \kk\cdot
\frac{T}{m}\ll 1\,.
\eqlabel{hydroexp1}
\end{equation}
The finite curvature $\frac{K}{m^2}\ne 0$ does not invalidate the
hydrodynamics, but simply restricts its applicability according to
\eqref{hydroexp1}. Indeed, compactification of the theory on $S^3$
'quantizes' the available spatial momenta according to \eqref{ks3}
\begin{equation}
\kk^2=\frac{K \ell (\ell+2)}{4\pi^2 T^2 }\,,\qquad \ell\in\zet_+\,,
\eqlabel{hext2}
\end{equation}
which constrains the order of the harmonic $\ell$ well described
by \eqref{sound}.  For example, for $K\ll \min\{T^2,m^2\}$,
\begin{equation}
2\le\  \ell\bigg|_{hydro}\ \ll\ \ell_{max}\equiv \min\left\{\frac{2\pi T}{\sqrt{K}}\,,\, \frac{2\pi m}{\sqrt{K}}\right\} \,.
\eqlabel{hext3}
\end{equation}
Note the lower bound on $\ell$: the hydrodynamic
fluctuation are fluctuations of the energy density/pressure, which are
physical on $S^3$ only for $\ell\ge 2$, see appendix \ref{eomshol}.
Of course, one can improve the agreement of the exact QNM dispersion
$\ww(\kk)$ with its hydrodynamic approximation \eqref{sound}  by  
including the higher-order hydrodynamic transport coefficients,
specifically the second-order curvature coupling coefficient $\kappa$
\cite{Baier:2007ix}. In this paper we compute $\ww(\kk)$
for generic $\kk$, without relying on hydrodynamic approximations:
$\ell=0$ and $\ell=1$ instabilities (see fig.~\ref{l0and1}) are always outside
the hydrodynamic approximation; likewise are outside the hydrodynamic
approximation  all $\ell=2$ branches in fig.~\ref{l2} except
for the black and the brown subbranches for $\frac{K}{m^2}\ll 1$.

\begin{itemize}
\item As discussed in appendix \ref{eomshol}, QNMs of the
black holes with spherical horizon at $\ell=\{0,1\}$ are physically distinct
from those with $\ell\ge 2$: the former ones include physical
fluctuations exclusively in the bulk gravitational scalar sector; while the
latter ones mix the bulk scalar and the physical metric fluctuations.
In fig.~\ref{l0and1} we present the dispersion of $\ell=0$ (solid curves)
and $\ell=1$ (dashed curves) QNMs with $\frac{K}{m^2}$ at fixed
temperature  \eqref{tm}. At $K=0$ these modes start from 
the magenta QNMs of fig.~\ref{f0mode1K0}.
The $\ell=1$ mode is stable. 
While the $\ell=0$ mode starts as the stable ones at $K=0$, it develops an
instability in the range
\begin{equation}
\frac{K_{\ell=0}^{stable}}{m^2}\leq \frac{K}{m^2}\leq
\frac{K_{\ell=0}^{unstable}}{m^2}\,.
\eqlabel{l0insta}
\end{equation}
Of course, the starting point
at $K=0$ can be {\it any} AdS$ _5$ QNM of fig.~\ref{adsqnm} evolved
to the required value of $\frac{m^2}{T_p^2}$, see  \eqref{tm} --- \eg
in fig.~\ref{l0exc} we include the $\frac{K}{m^2}$ dispersion
of the $\ell=0$ and $\ell=1$ modes starting from the pink QNMs of fig.~\ref{f0mode2K0}. 
\item In fig.~\ref{l2} we present the dispersion of $\ell=2$  QNMs with $\frac{K}{m^2}$ at fixed temperature  \eqref{tm}. We discuss only the modes that
originate at $K=0$ from the magenta QNMs of  fig.~\ref{f0mode1K0},
and the hydrodynamic (sound) mode (the cyan dot with $\Im[\ww]=0$).
The $\ell=2$ mode  remains unstable from $K=0$ up to $K_{\ell=2}^{unstable}$,
\begin{equation}
0\leq \frac{K}{m^2}\leq \frac{K_{\ell=2}^{unstable}}{m^2}\,.
\eqlabel{l2insta}
\end{equation}
\item Interestingly, $K_{\ell=2}^{unstable}<K_{\ell=0}^{stable}$, thus
there is stability range of $\caln=2^*$ black holes at the intermediate values
of $K$:
\begin{equation}
\frac{K_{\ell=2}^{unstable}}{m^2}< \frac{K}{m^2}< \frac{K_{\ell=0}^{stable}}{m^2}\,.
\eqlabel{lstable}
\end{equation}
\item Finally, in fig.~\ref{higherl} we present the dispersion of
higher-$\ell$ QNMs with $\frac{K}{m^2}$ at fixed temperature  \eqref{tm}.
We highlight only the unstable subbranch of the QNMs, the one that
originates at $K=0$ from the hydrodynamic (sound) mode. All these modes
are unstable similar to $\ell=2$ mode, see \eqref{l2insta}, with
distinct $K_\ell^{unstable}$ such that
\begin{equation}
K_\ell^{unstable}\ >\ K_{\ell'}^{unstable}\,,\qquad {\rm if}\qquad \ell'\ >\ \ell\,.
\eqlabel{lhierarhy}
\end{equation}
\end{itemize}

In the following plots we always present dimensionless quantities.
Dimensionless frequency and momenta of the QNMs are defined as in footnote 1. 
We further introduce the reduced free energy density $\hat\calf$, the reduced
energy density $\hat\cale$, and the reduced entropy density $\hat s$ as follows:
\begin{equation}
\begin{split}
\hat\calf=\frac{8}{3\pi^2N^2}\ \frac{\calf}{m^4}\,,\qquad
\hat\cale=\frac{8}{3\pi^2N^2}\ \frac{\cale}{m^4}\,,\qquad \hat s=\frac{8}{3\pi^2N^2}\
\frac{s}{m^3}\,,
\end{split}
\eqlabel{defhat}
\end{equation}
where the overall prefactor is chosen in such a way that
\begin{equation}
\lim_{m\to 0}\ \biggl[\frac{m^4}{T^4}\ \hat\cale\biggr]=1 \,.
\eqlabel{adslimmit}
\end{equation}

\subsection{\underline{(A)}}

\begin{figure}[t]
\begin{center}
\psfrag{m}[cc][][1][0]{$\frac{m^2}{T^2}$}
\psfrag{f}[bb][][1][0]{$\hat\calf$}
\psfrag{d}[tt][][0.7][0]{$({\color{green}\delta\hat\calf})^{2/3}\equiv
({\color{red}\hat\calf}-{\color{blue}\hat\calf})^{2/3}
$}
\psfrag{e}[cc][][1][0]{$\hat\cale$}
\psfrag{s}[bb][][1][0]{$\hat s$}
\includegraphics[width=3in]{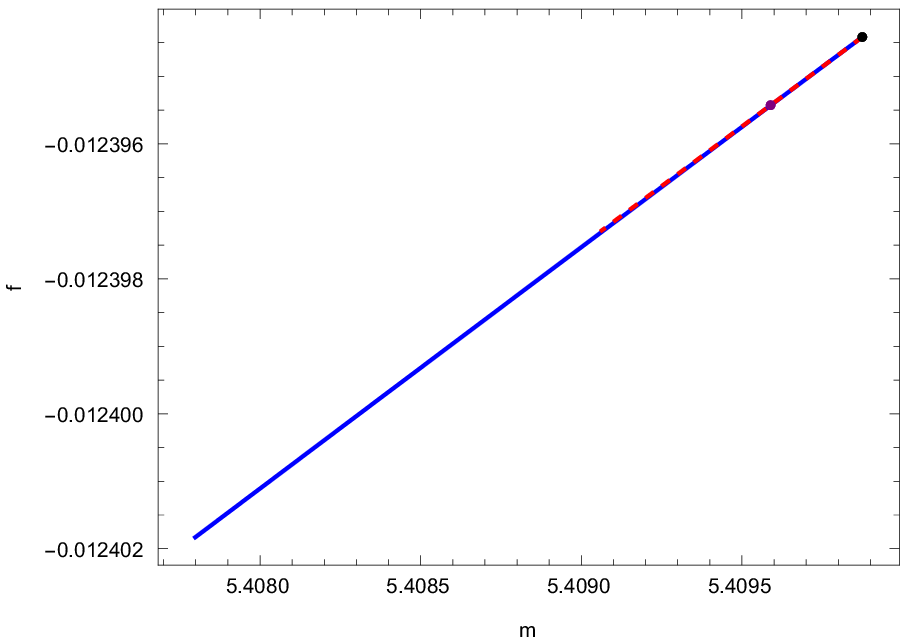}
\includegraphics[width=3in]{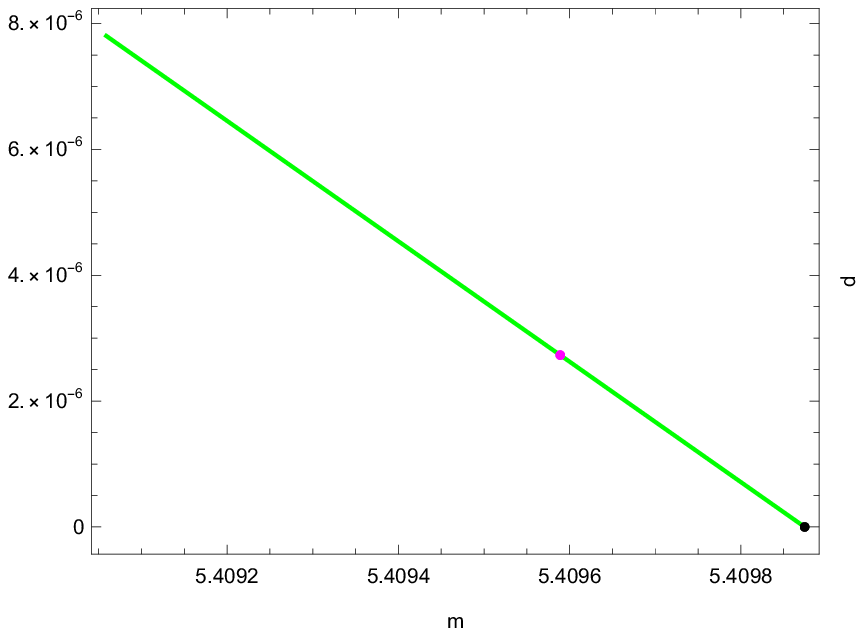}
\end{center}
  \caption{Thermodynamics of $\caln=2^*$ model in the canonical ensemble at $K=0$.
  The solid blue curve is the thermodynamically stable phase with $c_v>0$;
  the dashed red curve is the thermodynamically unstable phase with
  $c_v<0$. Both phases join at the terminal temperature, denoted by the black dot.
  The stable phase has a lower free energy density, and is
  the thermodynamically preferred one (right panel). The purple dot indicates
  the thermal state (in the thermodynamically unstable phase) were we compute
  the QNM spectra and follow them as $K\ne 0$.
} \label{macro}
\end{figure}

In fig.~\ref{macro}  we present the thermodynamics of the model
at $K=0$ in the canonical ensemble \cite{Buchel:2010ys}.
There is a critical (terminal) temperature 
$T_{crit,0}$ (denoted by the black dot)
\begin{equation}
\frac{m^2}{T_{crit,0}^2}=5.4098(7)\,,
\eqlabel{trcit0}
\end{equation}
where the two phases join: the solid blue curve is the phase with $c_v>0$ while
the red dashed curve represents the phase with $c_v<0$. As $T\to T_{crit,0}$
(from above) the specific heat diverges as 
\begin{equation}
\frac{c_v}{s}\ \propto \pm \left(1-\frac{T_{crit,0}}{T}\right)^{-1/2}\,.
\eqlabel{cvdiv}
\end{equation}
The free energy density of the thermodynamically unstable phase is always
above that of the stable phase. This is more clear in the right panel, were we
plot
\begin{equation}
({\color{green}\delta\hat\calf})^{2/3}\equiv
({\color{red}\hat\calf}-{\color{blue}\hat\calf})^{2/3}\ \propto
+\left(1-\frac{T_{crit,0}}{T}\right)\,,\qquad {\rm as}\qquad T\to T_{crit,0}\,.
\eqlabel{diff}
\end{equation}
The linear scaling of $(\delta\hat\calf)^{2/3}$ with the temperature near the
criticality is a reflection of \eqref{cvdiv}.
Note the purple dot in the red (thermodynamically unstable phase) --- more pronounced
in the right panel --- at temperature $T_p$,
\begin{equation}
\frac{m^2}{T_p^2}=5.40959\,.
\eqlabel{tm}
\end{equation}
We will study the QNM spectra of the model for different $\frac {K}{m^2}$,
keeping the temperature fixed as in \eqref{tm}.

\begin{figure}[t]
\begin{center}
\psfrag{m}[cc][][1][0]{$\frac{m^2}{T^2}$}
\psfrag{f}[bb][][1][0]{$\hat\calf$}
\psfrag{d}[tt][][0.7][0]{$({\color{green}\delta\hat\calf})^{2/3}\equiv
({\color{red}\hat\calf}-{\color{blue}\hat\calf})^{2/3}
$}
\psfrag{e}[cc][][1][0]{$\hat\cale$}
\psfrag{s}[bb][][1][0]{$\hat s$}
\includegraphics[width=3in]{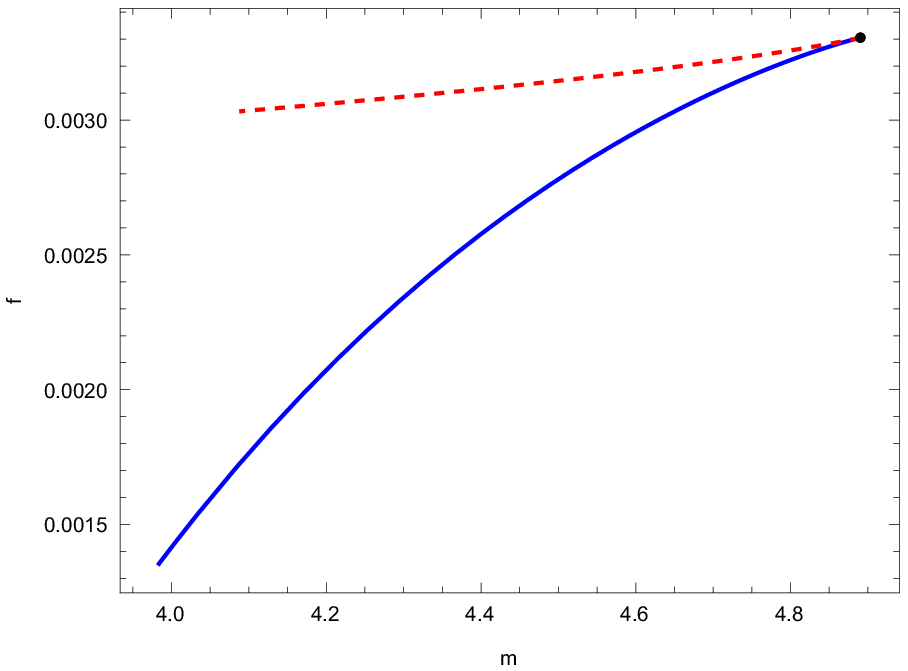}
\includegraphics[width=3in]{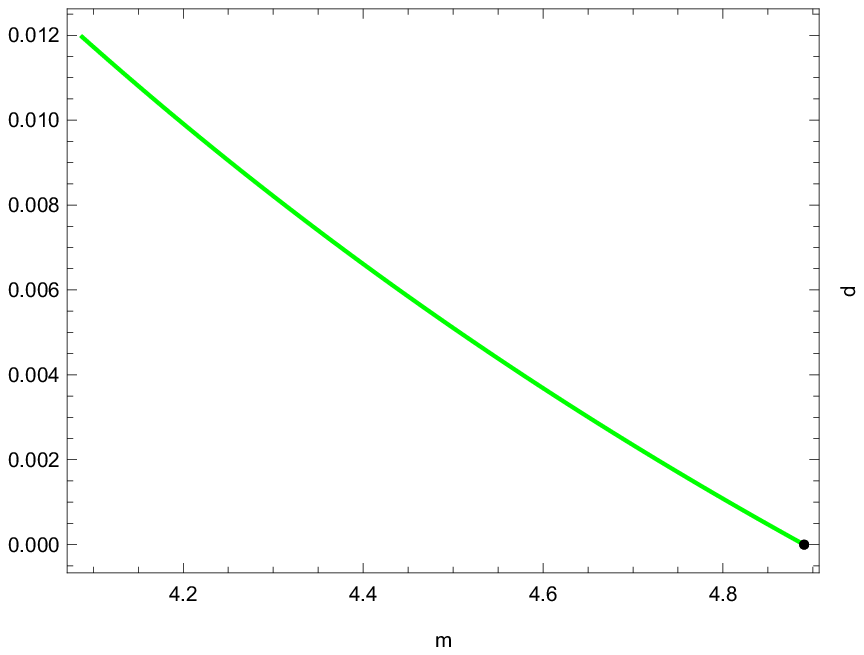}
\end{center}
  \caption{Thermodynamics of $\caln=2^*$ model in the canonical ensemble at $K=m^2$
  is qualitatively the same as for $K=0$, see fig.~\ref{macro}. 
} \label{macro1}
\end{figure}

As shown in fig.~\ref{macro1}, the canonical ensemble phase diagram is qualitatively
unchanged as $K\ne0$; here $K=m^2$. While the critical temperature at $\frac{K}{m^2}=1$,
$T_{crit,1}$ 
\begin{equation}
T_{crit,1}  > T_{crit,0}\,,
\eqlabel{tc10}
\end{equation}
we find that $T_{crit}$ is not a monotonically increasing function of $\frac{K}{m^2}$, \eg
\begin{equation}
\frac{m^2}{T_{crit}}\bigg|_{K/m^2=0.628169}=7.266(0)\ >\ \frac{m^2}{T_{crit}}\bigg|_{K/m^2=0}\ >\
\frac{m^2}{T_{crit}}\bigg|_{K/m^2=1}=4.8903(1)\,.
\eqlabel{628}
\end{equation}

\begin{figure}[t]
\begin{center}
\psfrag{m}[cc][][1][0]{$\frac{m^2}{T^2}$}
\psfrag{f}[cc][][1][0]{$\hat\calf$}
\psfrag{e}[cc][][1][0]{$\hat\cale$}
\psfrag{s}[bb][][1][0]{$\hat s$}
\psfrag{b}[tt][][1][0]{$\hat s$}
\includegraphics[width=3in]{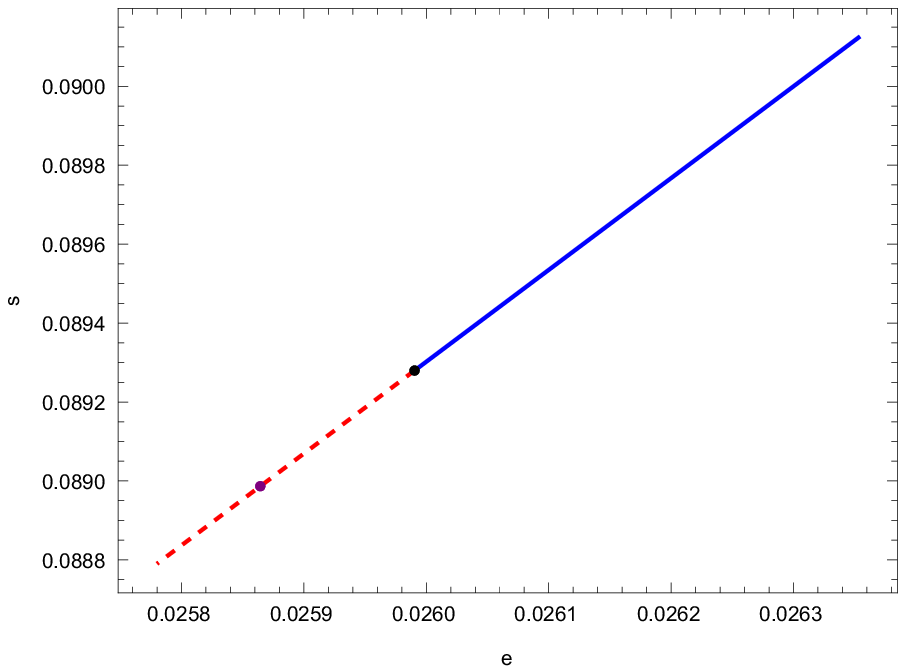}
\includegraphics[width=3in]{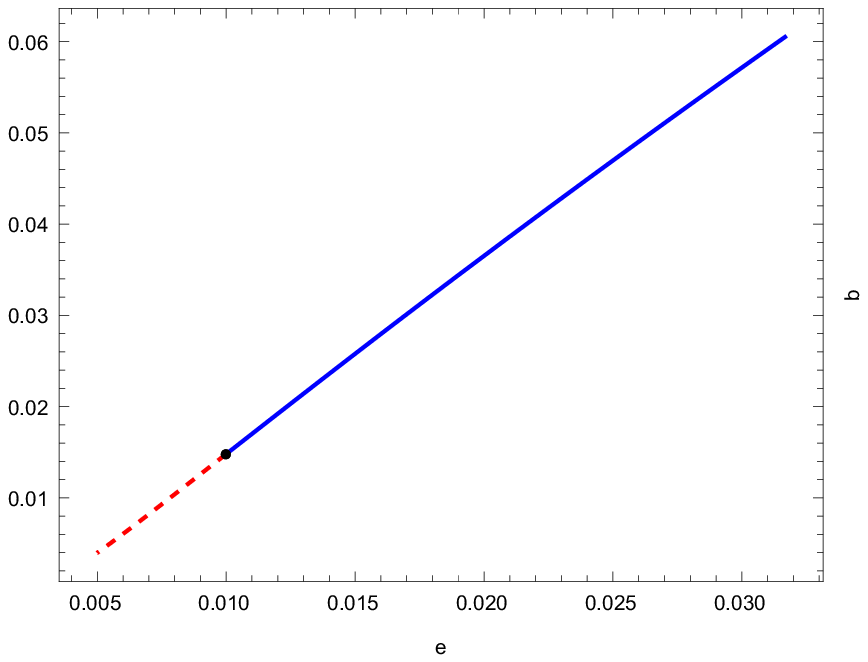}
\end{center}
  \caption{Thermodynamics of $\caln=2^*$ model in the microcanonical ensemble
  at $K=0$ (the left panel) and $K=m^2$ (the right panel).
  The thermodynamically stable phases (the solid blue curves) terminate
  at the $\hat\cale_{cit}$, denoted by the black dot.
Thermal states with $\hat\cale<\hat\cale_{crit}$ are thermodynamically
unstable (the red dashed curves).
The purple
  dot indicates the reduced energy density $\hat\cale_p$ of the state
  selected for the QNM spectra analysis. 
} \label{micro}
\end{figure}

In fig.~\ref{micro} we present the thermodynamics of the
model in the microcanonical ensemble: $K=0$ (the left panel) and
$K=m^2$ (the right panel). The black dots indicate the critical
point of the canonical ensemble at $\frac{K}{m^2}=0$ and $\frac{K}{m^2}=1$,
\begin{equation}
\hat\cale_{crit,0}=0.025990(4)\,,\qquad \hat\cale_{crit,1}=0.0099855(1)\,.
\eqlabel{hatcrit}
\end{equation}
The color coding is the same as in
figs.~\ref{macro} and \ref{macro1}. The purple dot (the left panel)
identifies the state which we use to study the QNM spectra of the
$\caln=2^*$ thermodynamically unstable horizon,
\begin{equation}
\hat\cale_{p}=0.0258644\,.
\eqlabel{hatem}
\end{equation}

\begin{figure}[t]
\begin{center}
\psfrag{m}[cc][][1][0]{$\frac{m^2}{T^2}$}
\psfrag{c}[bb][][1][0]{$c_s^2$}
\psfrag{z}[tt][][1][0]{$\zeta/\eta$}
\includegraphics[width=3in]{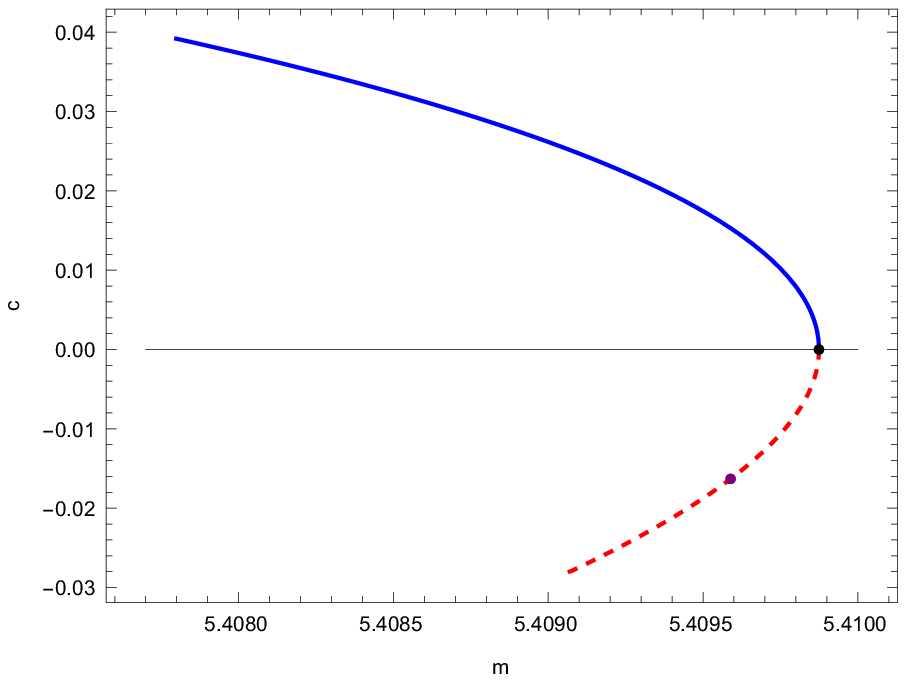}
\includegraphics[width=3in]{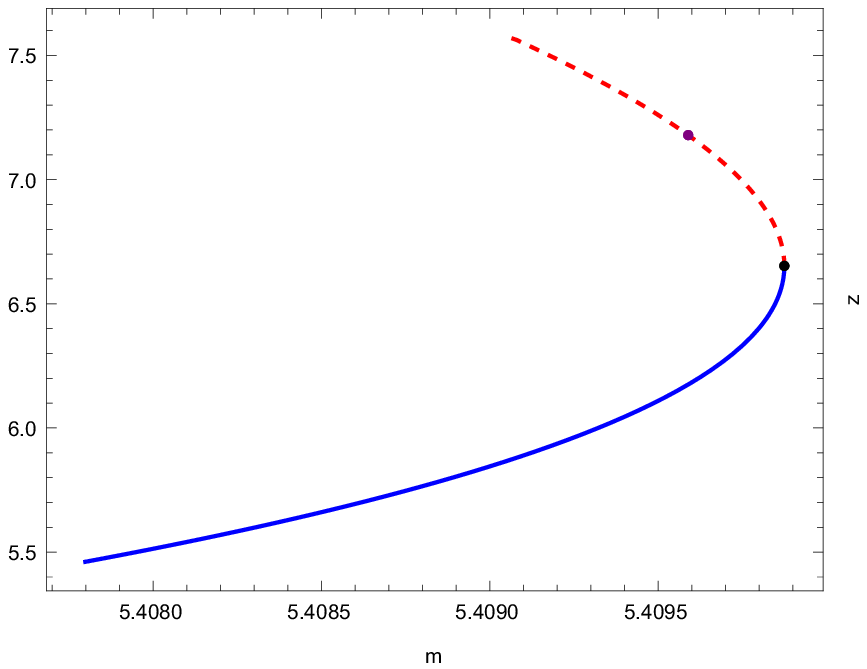}
\end{center}
  \caption{The speed of the sound waves $c_s^2$ (the left panel) and
  the ratio of bulk-to-shear viscosities $\frac{\zeta}{\eta}$
  (the right panel)
  in $\caln=2^*$ black branes close to a critical
  point \eqref{trcit0} (the black dot). Thermodynamically stable
  and unstable phases are represented by the solid and the  dashed
  curves correspondingly. The purple dot identifies
  the thermal state of interest, see \eqref{transport} for the
  values of its transport coefficients. 
} \label{csbv}
\end{figure}

In fig.~\ref{csbv} we present the results for the speed of the
sound waves \cite{Buchel:2007vy} (the left panel),
and the bulk viscosity (the right panel) of the $\caln=2^*$ black
branes \cite{Buchel:2007mf,Buchel:2008uu,Buchel:2011yv}. For consistency,
we use the same color coding as in earlier plots: the solid blue curves
denote the thermodynamically stable phase, and the red dashed ones
represent the thermodynamically unstable phase. As expected from
\cite{Buchel:2005nt}, $c_s^2<0$ in the thermodynamically unstable phase.
Note that the speed of the sound waves vanishes at $T_{crit,0}$, see
\eqref{trcit0}, responsible (according to \eqref{cscv}) for the
divergence of the specific heat \eqref{cvdiv}. The bulk viscosity
is finite at the criticality, denoted as the black dot.
The unstable thermal state of interest once again is indicated with
the purple dot ---
here the temperature $T_p$ is given by \eqref{tm}, and the specific values
of the transport coefficients are
\begin{equation}
c_s^2=-0.016312(8)\,,\qquad \frac{\zeta}{\eta}=7.1791(6)\,,\qquad \frac{\eta}{s}=\frac{1}{4\pi}\,,
\eqlabel{transport}
\end{equation}
where we included the universal result for the shear viscosity $\eta$ as
well \cite{Buchel:2003tz,Kovtun:2004de}.
The knowledge of the transport \eqref{transport} provides
a prediction for the dispersion of the hydrodynamic sound waves
as in \eqref{sound}, and a valuable test of the QNM computations
reported in fig.~\ref{hydroK0}
in the framework of appendix \ref{eomshol}.

\begin{figure}[t]
\begin{center}
\psfrag{q}[cc][][1][0]{$\kk$}
\psfrag{i}[bb][][1][0]{$\Im[\ww]$}
\psfrag{r}[tt][][1][0]{$\Re[\ww]$}
\includegraphics[width=3in]{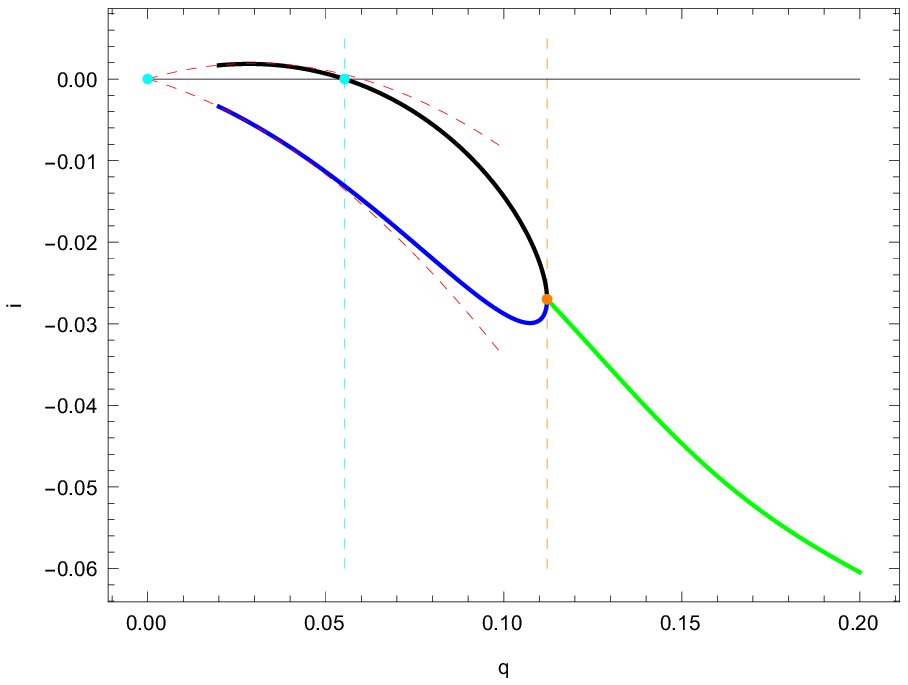}
\includegraphics[width=3in]{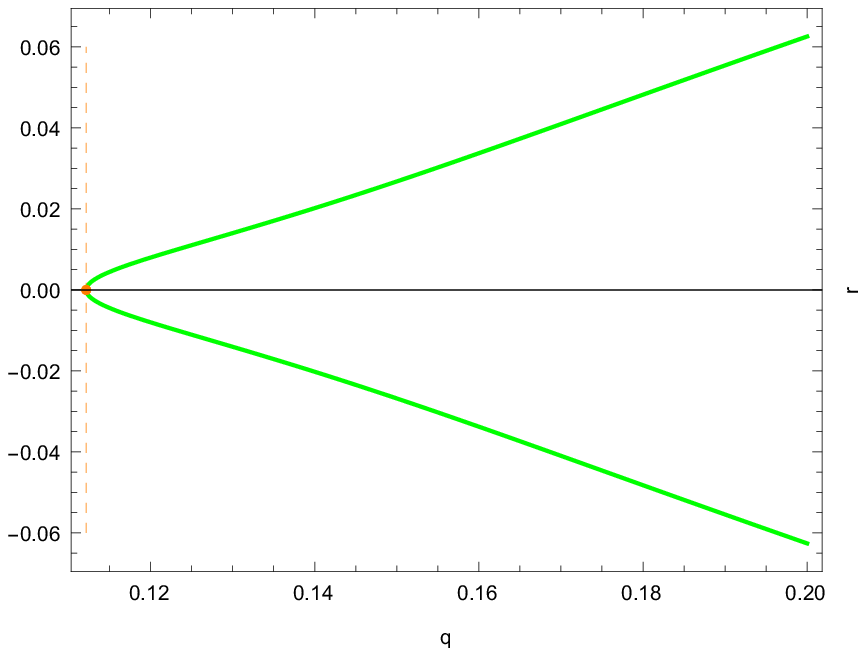}
\end{center}
  \caption{The dispersion relation of the hydrodynamic QNM $\ww(\kk)$
  of the thermodynamically unstable state of the $\caln=2^*$ black brane
  at temperature \eqref{tm}. The dashed red curves represent the
  $\calo(\kk^2)$ hydrodynamic approximation \eqref{sound}. The mode is
  nonpropagating for $\kk<\kk_o$, represented by the vertical dashed
  orange lines.  It is unstable (the black branch) in the range of $\kk$ between
  the two cyan dots, see \eqref{unstablek0}. 
} \label{hydroK0}
\end{figure}

In fig.~\ref{hydroK0} we present the dispersion $\ww=\ww(\kk)$ of the
hydrodynamic quasinormal mode of the $\caln=2^*$ black branes
in the thermodynamically unstable state represented by
the purple dot in figs.~\ref{macro},\ref{micro} and \ref{csbv}, at temperature
\eqref{tm}, and the energy density \eqref{hatem}. 
The dashed red curves represent the $\calo(\kk^2)$ hydrodynamic
approximation \eqref{sound},
with the appropriate transport coefficients \eqref{transport}.
The solid black and and blue curves represent $\ww(\kk)$
for the unstable and the stable subbranches of the hydrodynamic QNM
computed in the framework of appendix \ref{eomshol} for $\kk\ge \frac{1}{50}$.
Note the agreement  with the hydrodynamic predictions \eqref{sound}
for small $\kk$. The black/blue subbranches are purely imaginary,
and coalesce at
\begin{equation}
\kk_o=0.11215(1)\,,
\eqlabel{kor}
\end{equation}
represented by the vertical dashed orange lines. For $\kk>\kk_o$ the
hydrodynamic sound mode, non-propagating for $\kk\in [0,\kk_o]$, becomes
propagating, \ie it develops $\Re[\ww]\ne 0$. The propagating part of the
sound hydrodynamic QNM is represented with solid green curves.
As expected from the general arguments \cite{Buchel:2005nt}, the $\caln=2^*$
thermodynamically unstable black brane state \eqref{tm}  is dynamically unstable for 
\begin{equation}
\kk\in (0,\kk_{unstable})\,,\qquad \kk_{unstable}=0.055273(8)\,,
\eqlabel{unstablek0}
\end{equation}
represented by the vertical dashed cyan
line.  

\subsection{\underline{(B)}}

\begin{figure}[t]
\begin{center}
\psfrag{i}[bb][][1][0]{$\Im[\ww]$}
\psfrag{b}[tt][][1][0]{$\Im[\ww]$}
\psfrag{r}[cc][][1][0]{$\Re[\ww]$}
\includegraphics[width=3in]{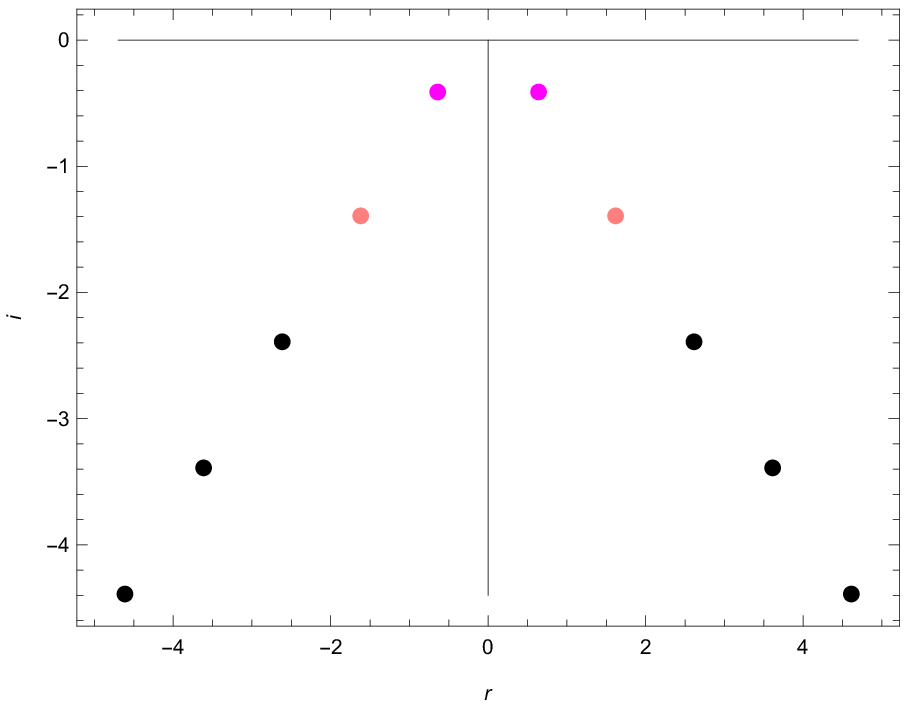}
\end{center}
  \caption{Non-hydrodynamic QNMs of $\caln=2^*$ black brane in the
  limit $\frac{m^2}{T^2}\to 0$ at $\kk=0$.  
} \label{adsqnm}
\end{figure}

\begin{figure}[t]
\begin{center}
\psfrag{m}[cc][][1][0]{$\frac{m^2}{T^2}$}
\psfrag{i}[bb][][1][0]{$\Im[\ww]$}
\psfrag{r}[tt][][1][0]{$\Re[\ww]$}
\includegraphics[width=3in]{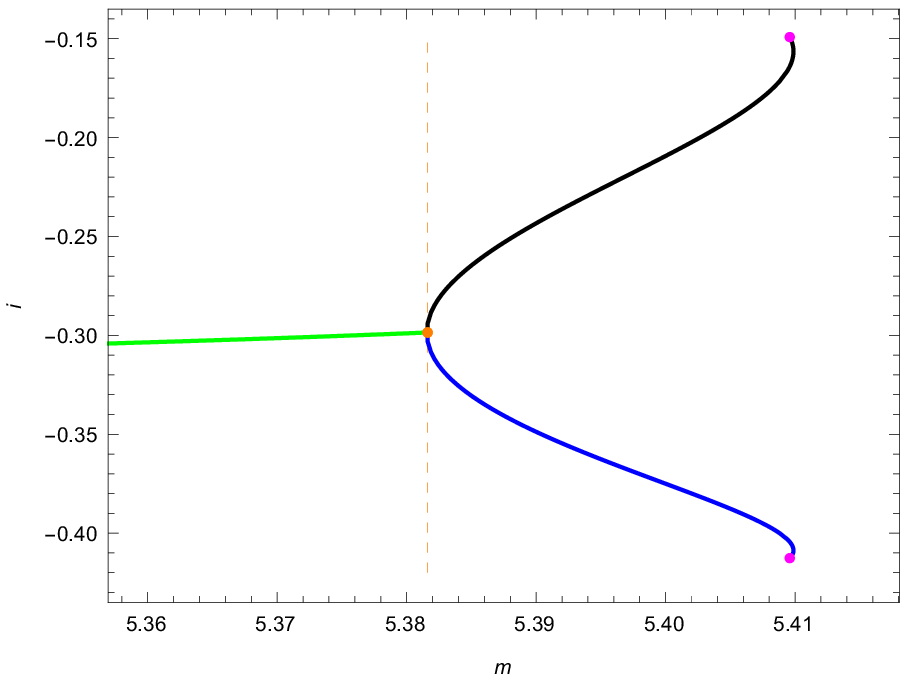}
\includegraphics[width=3in]{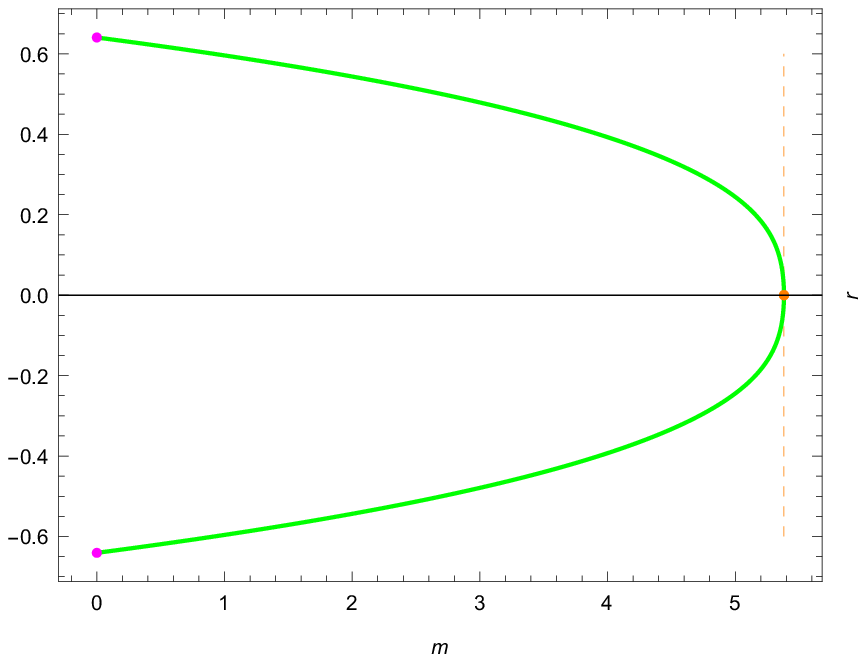}
\end{center}
  \caption{We follow the pair of magenta QNMs of fig.~\ref{adsqnm},
  at $\kk=0$,
  increasing $m$ to reach the thermodynamically unstable $\caln=2^*$
  black brane state \eqref{tm}. For $T_p\leq T< T_o$  these modes
are purely dissipative. The temperature $T_o$ is represented by
vertical dashed orange lines. 
} \label{f0mode1K0}
\end{figure}

\begin{figure}[t]
\begin{center}
\psfrag{m}[cc][][1][0]{$\frac{m^2}{T^2}$}
\psfrag{i}[bb][][1][0]{$\Im[\ww]$}
\psfrag{r}[tt][][1][0]{$\Re[\ww]$}
\includegraphics[width=3in]{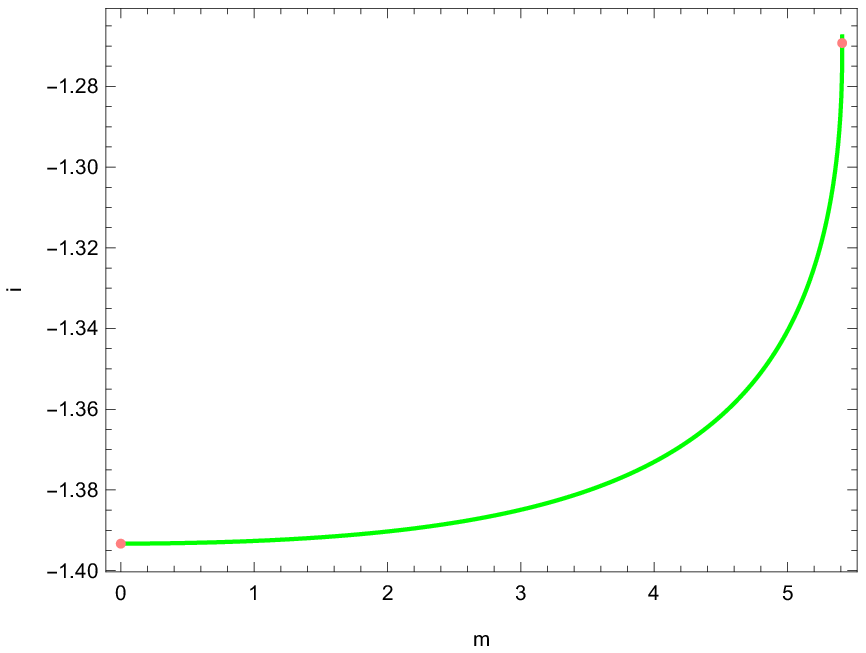}
\includegraphics[width=3in]{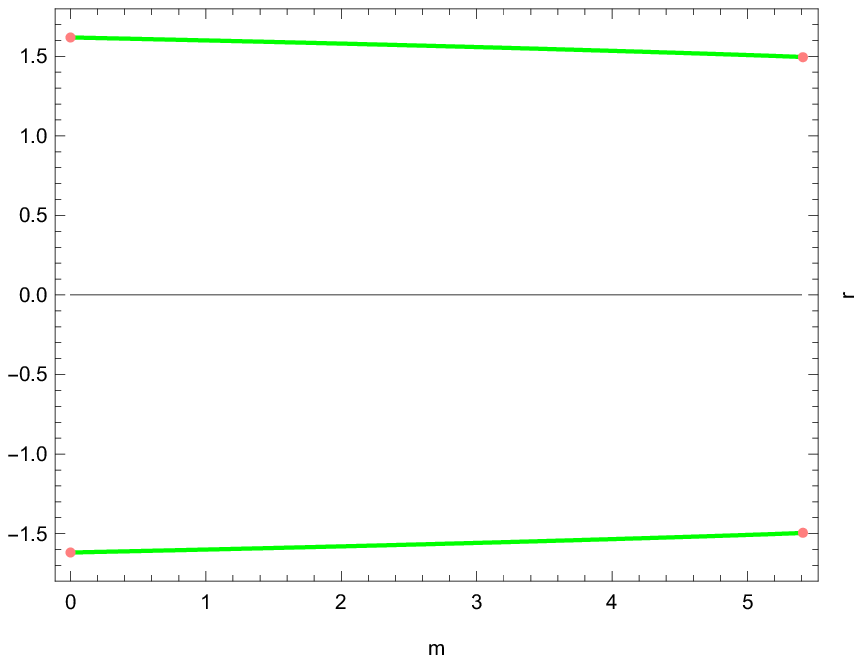}
\end{center}
  \caption{We follow the pair of pink QNMs of fig.~\ref{adsqnm},
  at $\kk=0$,
  increasing $m$ to reach the thermodynamically unstable $\caln=2^*$
  black brane state \eqref{tm}. These modes remain stable, with
  $\Re[\ww]\ne 0$. 
} \label{f0mode2K0}
\end{figure}

We begin discussion of the non-hydrodynamic QNMs of the
$\caln=2^*$ black brane at $m=0$, \ie in the pure
AdS$ _5$-Schwarzschild limit in fig.~\ref{adsqnm}.
There are two distinct $h=0$ subsectors, which are decoupled
when $m=0$. One of these subsectors contains a hydrodynamic
sound mode; we focus on the other subsector at $\kk=0$.
We highlight the two lowest pairs of the QNMs: the magenta and the pink one.
In fig.~\ref{f0mode1K0} we follow the magenta QNMs, increasing $m$ to reach
the thermodynamically unstable state of interest at temperature \eqref{tm}:
\begin{equation}
0\ \leq\ \frac{m^2}{T^2}\ \leq\  \frac{m^2}{T_p^2}\,.
\eqlabel{followm}
\end{equation}
While these QNMs start having both the real and the imaginary part of
the frequency $\ww$ at $m=0$, they become purely
dissipative for $T<T_o$,
\begin{equation}
\frac{m^2}{T_o^2}=5.38162\ <\ \frac{m^2}{T_p^2}\,,
\eqlabel{deftempo}
\end{equation}
represented by a vertical dashed orange line. 
Note that these modes are stable in the thermodynamically unstable
$\caln=2^*$ black brane state of interest. In fig.~\ref{f0mode2K0}
we follow in the same fashion the pink QNMs of fig.~\ref{adsqnm}:
they remain stable, and have $\Re[\ww]\ne 0$ up to the temperature
\eqref{tm}.

\subsection{\underline{(C)}}

We fix the temperature at \eqref{tm}, \ie in the thermodynamically
unstable state of the (originally) $\caln=2^*$ black brane, and study the
QNMs as the horizon curvature $K$ becomes nonzero.

\begin{figure}[t]
\begin{center}
\psfrag{k}[cc][][1][0]{$\frac{K}{m^2}$}
\psfrag{w}[bb][][1][0]{$\Im[\ww]$}
\includegraphics[width=5in]{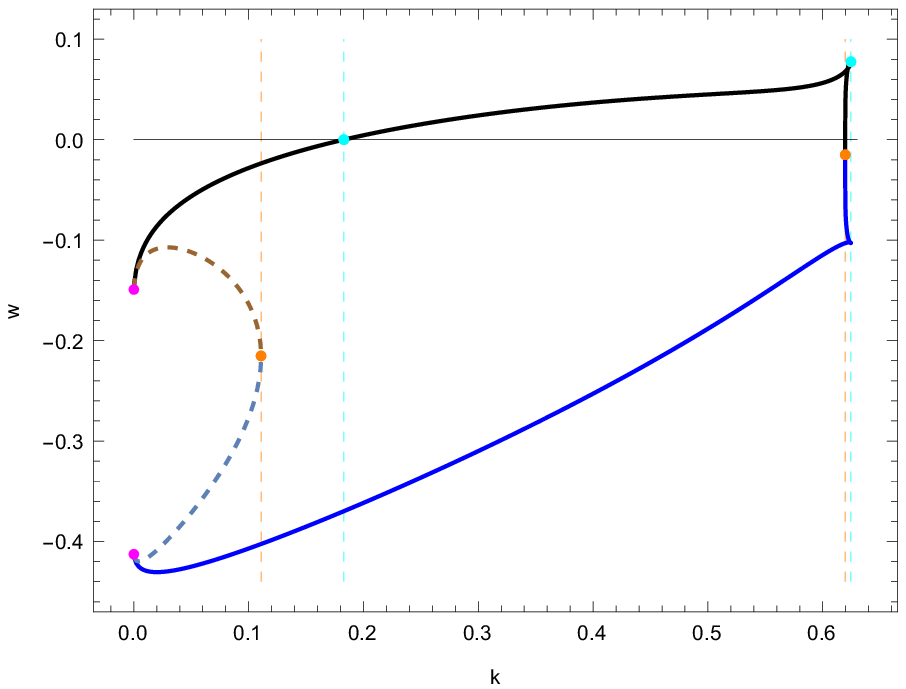}
\end{center}
  \caption{$\ell=0$ (solid curves) and $\ell=1$ (dashed curves)
  QNMs of $\caln=2^*$ black holes at fixed temperature \eqref{tm} as functions
  of the curvature $K$. The $\ell=0$ (solid black) subbranch is
  unstable in the range of $K$ between the vertical dashed cyan lines.
  QNMs at $K=0$ originate from the (different) magenta modes of
  fig.~\ref{f0mode1K0} --- they coalesce as stable QNMs,
  represented by the orange dots.
} \label{l0and1}
\end{figure}

\begin{figure}[t]
\begin{center}
\psfrag{k}[cc][][1][0]{$\frac{K}{m^2}$}
\psfrag{w}[bb][][1][0]{$\Im[\ww]$}
\psfrag{b}[tt][][1][0]{$\Im[\ww]$}
\includegraphics[width=3in]{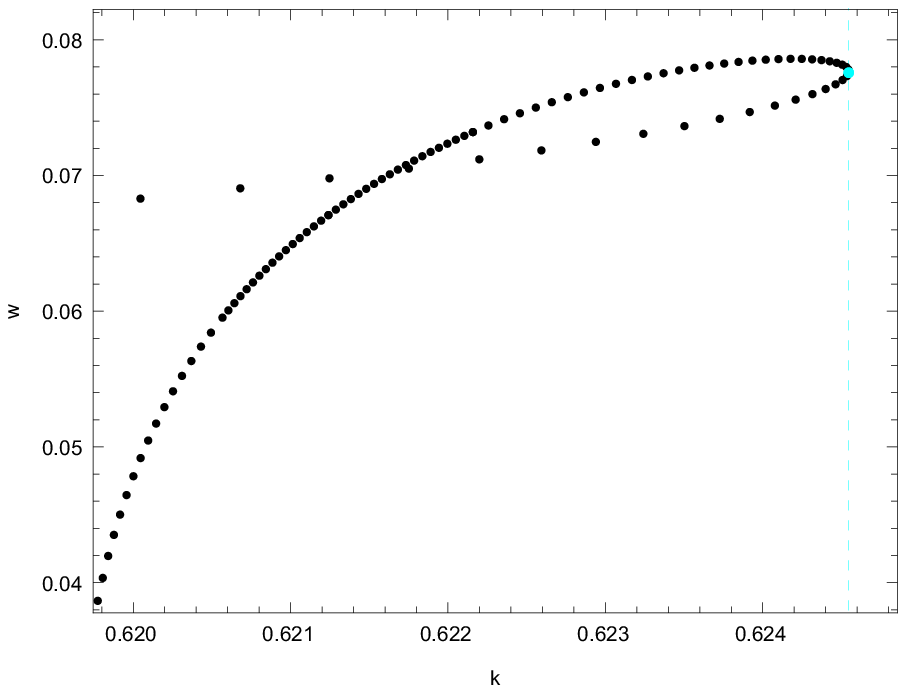}
\includegraphics[width=3in]{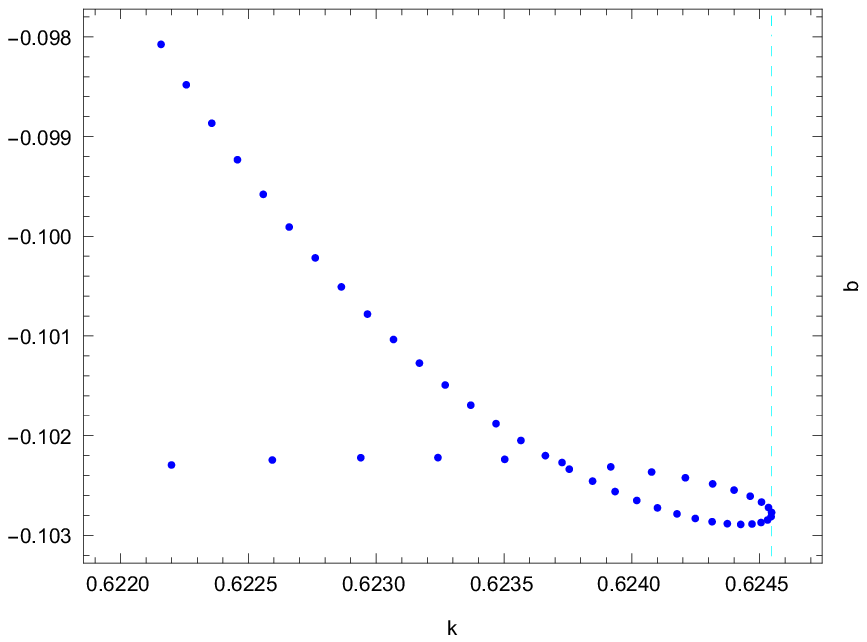}
\end{center}
  \caption{$\ell=0$ $\caln=2^*$
  black hole QNMs, originated  from the magenta modes of
  fig.~\ref{f0mode1K0}, are not single-valued
  functions of $\frac{K}{m^2}$
  in the vicinity of $K_{\ell=0}^{unstable}$, represented by
  the vertical dashed cyan lines.
} \label{l0topbot}
\end{figure}

\begin{figure}[t]
\begin{center}
\psfrag{k}[cc][][1][0]{$\frac{K}{m^2}$}
\psfrag{i}[bb][][1][0]{$\Im[\ww]$}
\psfrag{r}[tt][][1][0]{$\Re[\ww]$}
\includegraphics[width=3in]{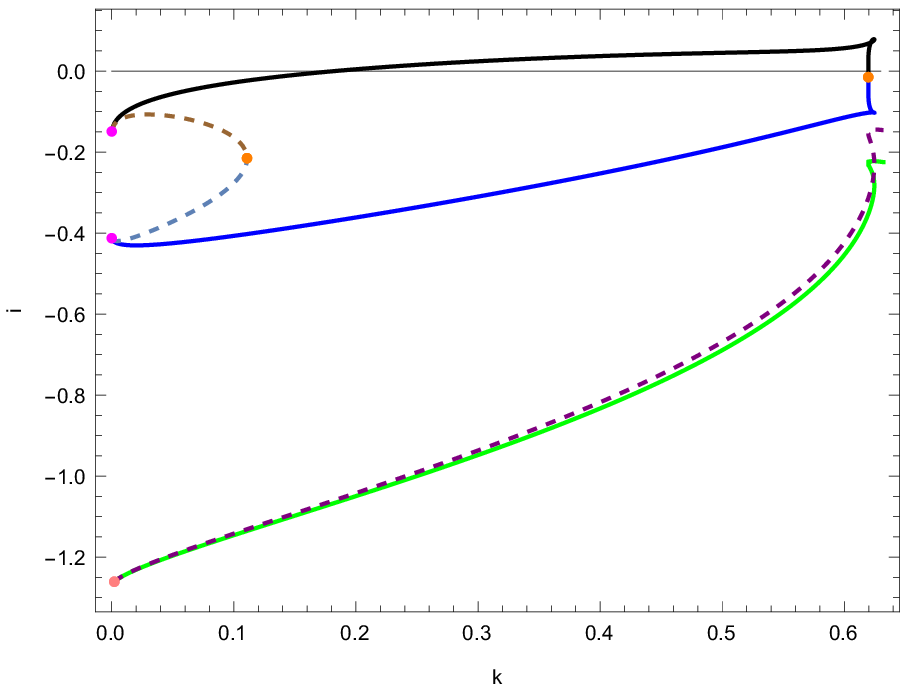}
\includegraphics[width=3in]{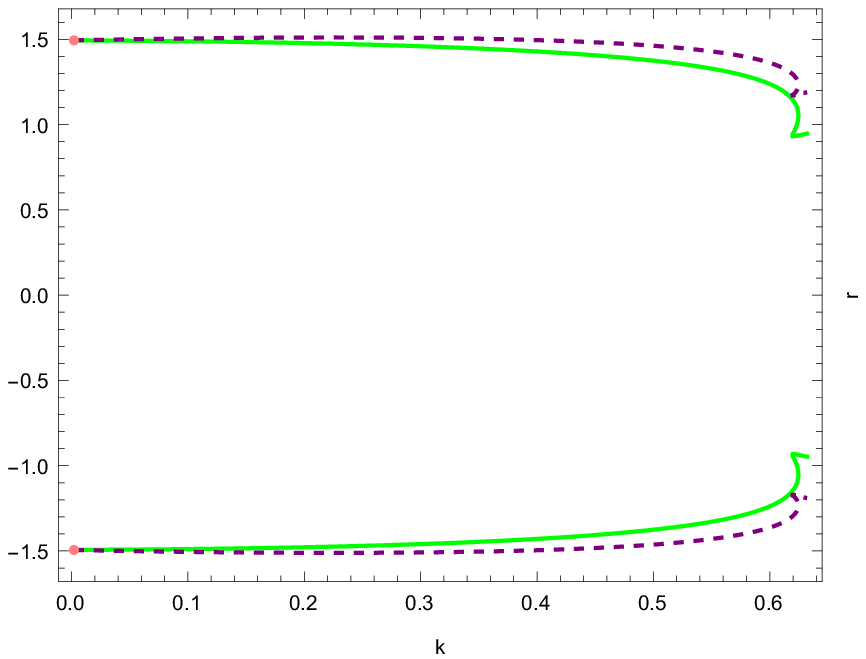}
\end{center}
  \caption{Stable
  $\ell=0$ (solid green curves) and $\ell=1$ (dashed purple curves)
  QNMs of $\caln=2^*$ black holes at fixed temperature \eqref{tm} as functions
  of the curvature $K$, that originate from the pink QNMs at
  $K=0$, see fig.~\ref{f0mode2K0}.
} \label{l0exc}
\end{figure}

\begin{figure}[t]
\begin{center}
\psfrag{k}[cc][][1][0]{$\frac{K}{m^2}$}
\psfrag{i}[bb][][1][0]{$\Im[\ww]$}
\psfrag{r}[tt][][1][0]{$|\Re[\ww]|$}
\includegraphics[width=3in]{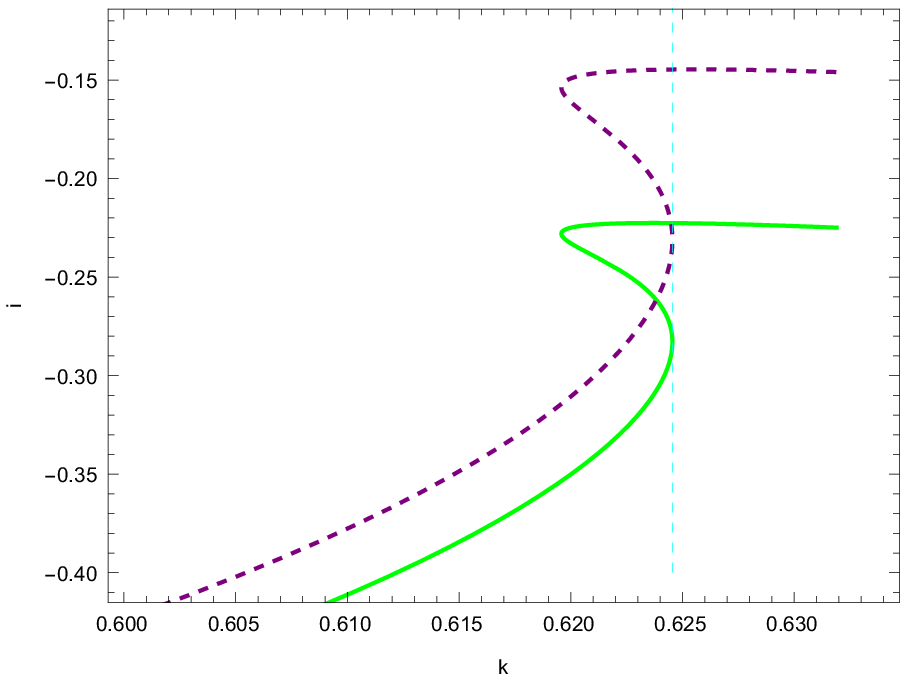}
\includegraphics[width=3in]{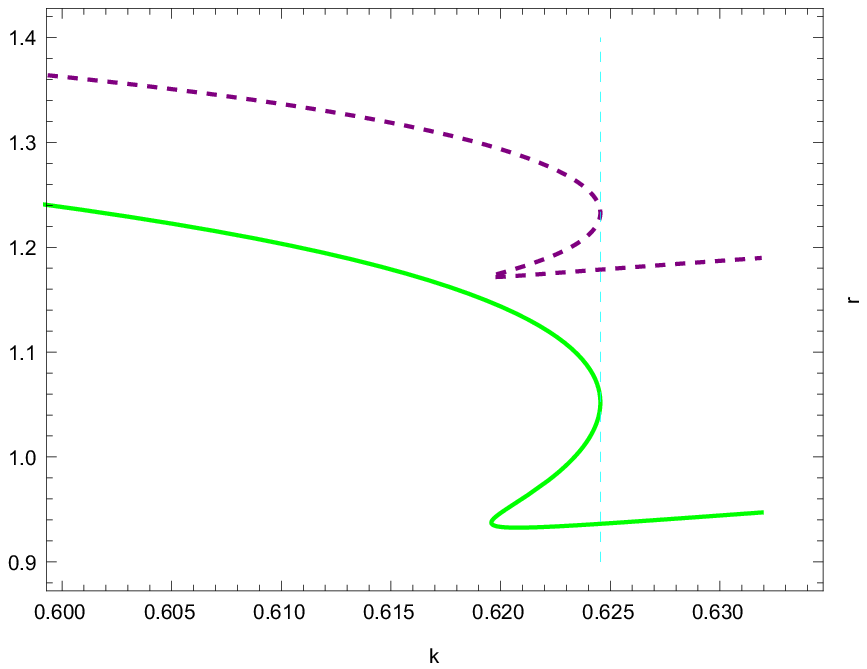}
\end{center}
  \caption{Both $\ell=0$ (solid green curves) and $\ell=1$
(dashed purple curves)
$\caln=2^*$
  black hole QNMs, originated  from the pink modes of
  fig.~\ref{f0mode2K0}, are not single-valued
  functions of $frac{K}{m^2}$
  in the vicinity of $K_{\ell=0}^{unstable}$, represented by
  the vertical dashed cyan lines.
} \label{notsingle}
\end{figure}

\begin{figure}[t]
\begin{center}
\psfrag{k}[cc][][1][0]{$\frac{K}{m^2}$}
\psfrag{i}[bb][][1][0]{$\Im[\ww]$}
\psfrag{r}[tt][][1][0]{$\Re[\ww]$}
\includegraphics[width=3in]{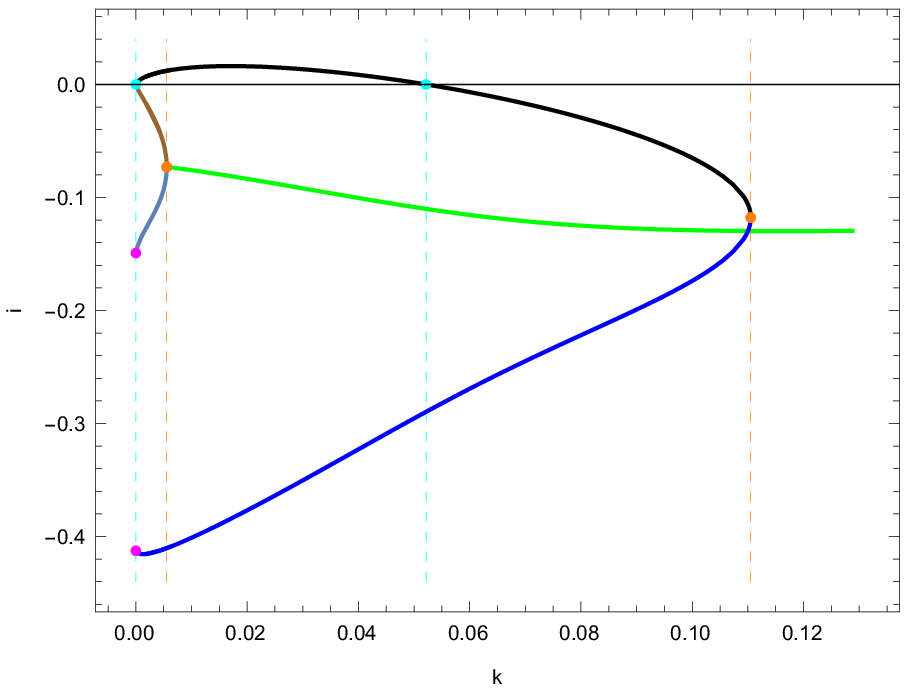}
\includegraphics[width=3in]{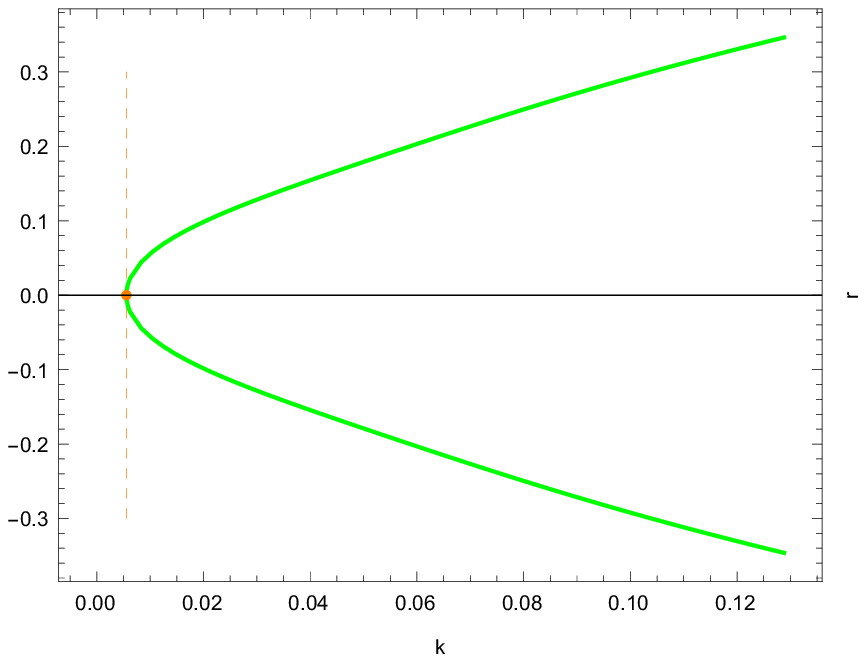}
\end{center}
  \caption{$\ell=2$
  QNMs of $\caln=2^*$ black holes at fixed temperature \eqref{tm} as functions
  of the curvature $K$. The solid black subbranch is
  unstable in the range of $K$ between the vertical dashed cyan lines.
  QNMs at $K=0$ originate from the  magenta modes of
  fig.~\ref{f0mode1K0} (grey and blue curves),
  and from the hydrodynamic QNM at $\ww=0$ (black and brown curves).
  The brown and grey subbranches coalesce, and continues as a
  green subbranch with $\Re[\ww]\ne 0$. No new subbranches originate once
  the black and the blue curves coalesce. 
} \label{l2}
\end{figure}

\begin{figure}[t]
\begin{center}
\psfrag{k}[cc][][1][0]{$\frac{K}{m^2}$}
\psfrag{i}[bb][][1][0]{$\Im[\ww]$}
\psfrag{b}[tt][][1][0]{$\Im[\ww]$}
\psfrag{x}[cc][][1][0]{$\ell$}
\includegraphics[width=3in]{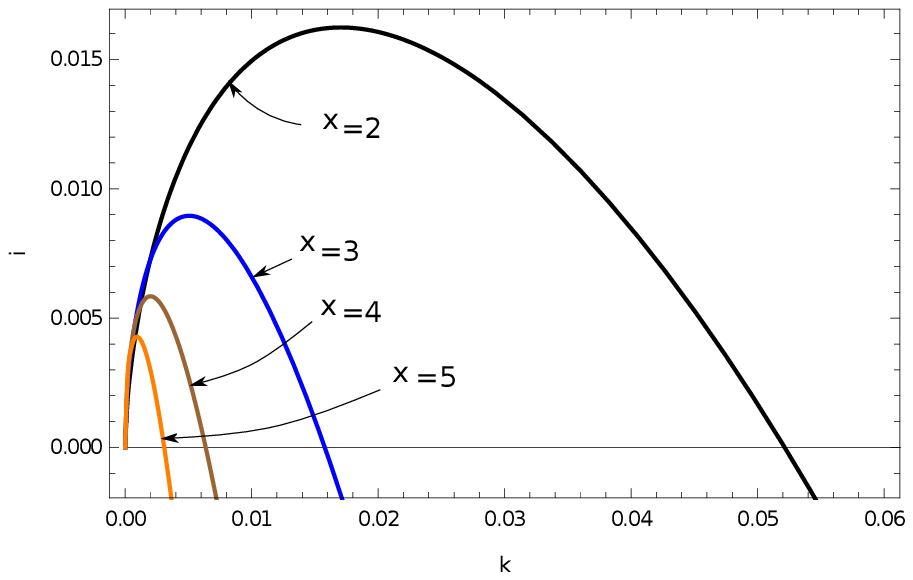}
\includegraphics[width=3in]{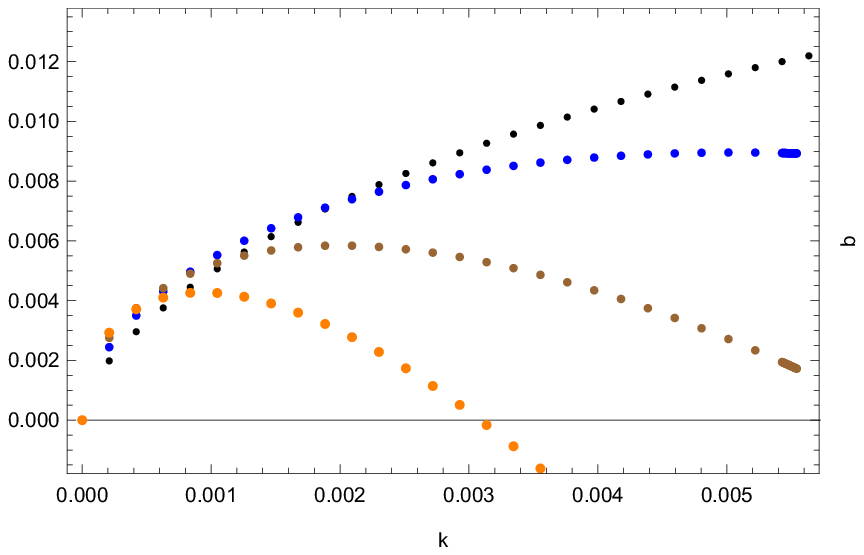}
\end{center}
  \caption{Subbranches of the $\ell=2,3,4,5$ QNMs of $\caln=2^*$ black holes
  at fixed temperature \eqref{tm} that contain instabilities. Note that
  higher-$\ell$ QNMs are stabilized at successively smaller values of
  $\frac{K}{m^2}$.
} \label{higherl}
\end{figure}

In fig.~\ref{l0and1} we present the dispersion of the
$\ell=0$ (solid curves) and $\ell=1$ (dashed curves) modes of
the $\caln=2^*$ black hole
with $\frac{K}{m^2}$, that originate from the magenta QNMs 
at $K=0$, see fig.~\ref{f0mode1K0}. These modes
have $\Re[\ww]=0$.
Initially stable $\ell=0$ mode becomes unstable (solid black curve) when
\begin{equation}
\frac{K}{m^2}\ >\ \frac{K_{\ell=0}^{stable}}{m^2}=0.18290(4)\,,
\eqlabel{cyanl0}
\end{equation}
represented by the vertical cyan line. The instability persists until
\begin{equation}
\frac{K}{m^2}\ <\ \frac{K_{\ell=0}^{unstable}}{m^2}=0.62454(7)\,,
\eqlabel{cyanl02}
\end{equation}
represented by the cyan vertical dashed line. 
The QNM subbranches originating from different QNMs at $K=0$
(the magenta dots) coalesce at 
\begin{equation}
\begin{split}
&\ell=1:\qquad \frac{K_{o,1}}{m^2}=0.11079(4)\,,\\
&\ell=0:\qquad \frac{K_{o,0}}{m^2}=0.61958(2)\,,
\end{split}
\eqlabel{orange01}
\end{equation}
as the stable QNMs, represented by the orange dots. Since
$K_{o,0}< K_{\ell=0}^{unstable}$, this implies that close to  $K_{\ell=0}^{unstable}$
the QNM subbranches can not be single-valued functions of $\frac{K}{m^2}$.
As fig.~\ref{l0topbot} shows this is indeed the case. See section \ref{tec}
for details how we are tracing the QNM curves. 

In fig.~\ref{l0exc} we present the dispersion of the
$\ell=0$ (solid green curves) and $\ell=1$ (dashed purple curves) modes of
the $\caln=2^*$ black hole
with $\frac{K}{m^2}$, that originate from the pink QNMs 
at $K=0$, see fig.~\ref{f0mode2K0}. These modes
have $\Re[\ww]\ne 0$, and are stable. We follow them to $K>K_{\ell=0}^{unstable}$
and observe that their $\Im[\ww]$ start decreasing again.
Close to $K_{\ell=0}^{unstable}$
these QNM subbranches are not single-valued functions of $\frac{K}{m^2}$ as well,
see fig.~\ref{notsingle}.

In fig.~\ref{l2} we present the dispersion of the $\ell=2$ modes
of the $\caln=2^*$ black holes with $\frac{K}{m^2}$, that originate
from the magenta QNMs at $K=0$ of fig.~\ref{f0mode1K0}
(grey and blue solid curves), and from the hydrodynamic QNM at $K=0$,
represented with the cyan dot at $\ww=0$ (black and brown solid curves).
The purely dissipative (stable) brown and grey branches (note that they
originate from the hydrodynamic (the cyan dot)
and the non-hydrodynamic (the magenta dot) modes at $K=0$)
coalesce at
\begin{equation}
\ell=2:\qquad \frac{K_{o,2,bg}}{m^2}=0.0055428(2)\,,
\eqlabel{kobg}
\end{equation}
represented by (the first) orange dot.
A new (stable) subbranch, represented by the solid green lines
starts of this orange dot.  
The QNM branch represented by the solid black curve contains the
instabilities seen at $K=0$ in fig.~\ref{hydroK0}: these QNMs are unstable in the range between the two cyan dashed vertical lines
\begin{equation}
0\ <\ \frac{K}{m^2}\ <\ \frac{K_{\ell=2}^{unstable}}{m^2}=0.0521(5)\,.
\eqlabel{l2unstable}
\end{equation}
The $\Re[\ww]=0$ black and  blue branches (note that they
originate from the hydrodynamic (the cyan dot)
and the non-hydrodynamic (the magenta dot) modes at $K=0$)
coalesce at
\begin{equation}
\ell=2:\qquad \frac{K_{o,2,bb}}{m^2}=0.11050(5)\,,
\eqlabel{kobb}
\end{equation}
represented by (the second) orange dot. No new subbranches
start at the second orange dot. Interestingly,
$K_{\ell=2}^{unstable}< K_{\ell=0}^{stable}$, as a result there is a finite range of
$\frac{K}{m^2}$, see \eqref{lstable}, where $\ell=2$ QNMs are already
stable, but $\ell=0$ QNMs have yet  to become unstable.  

We conclude the summary of the $\caln=2^*$ black hole QNMs at temperature
\eqref{tm} in fig.~\ref{higherl}: we show $\ell=\{2,3,4,5\}$ subbranches that originate from the hydrodynamic $\ww=0$ QNM at $K=0$, and contain
instabilities. All these modes are unstable as $\frac{K}{m^2}\to 0$,
and higher-$\ell$ modes are stabilized at successively smaller values
of $\frac{K}{m^2}$. Thus, the instabilities of $\ell> 2$ modes
do not further constrain the set of values of $\frac{K}{m^2}$
beyond that of the $\ell=2$ QNMs.

\section{Technical details}\label{tec}

The relevant effective action \cite{Pilch:2000fu} describing
$\caln=2^*$ black holes  is in the general class \eqref{efac} with $p=2$:
\begin{equation}
\begin{split}
&\phi_1\equiv \alpha\,,\qquad \eta_1=12\,;\qquad
\phi_2\equiv \chi\,,\qquad \eta_2=4\,,\\
&V=\frac14\ \biggl(\frac13\ \left(\frac{\del W}{\del\alpha}\right)^2
+\left(\frac{\del W}{\del\chi}\right)^2\biggr)-\frac43\ W^2\,,
\end{split}
\eqlabel{pw5d}
\end{equation}
where the superpotential $W$ is 
\begin{equation}
W=-e^{-2\alpha}-\frac 12\ e^{4\alpha}\ \cosh(2\chi)\,.
\eqlabel{pww}
\end{equation}
The scalars $\alpha$ and $\chi$ are the holographic dual to
operators $\calo_2$ and $\calo_3$ correspondingly, describing the
mass deformation of the maximally supersymmetric $\caln=4$
Yang-Mills\footnote{The field content of the $\caln=4$ SYM theory includes the
gauge field $A_\mu$, four Majorana fermions $\psi_a$ and three complex
scalars $\phi_i$, where all of these fields are in the adjoint
representation.} \cite{Buchel:2000cn,Buchel:2007vy,Hoyos:2011uh}
\begin{equation}
\begin{split}
&\call_{\caln=2^*}=\call_{\caln=4}-2 \left[ m_b^2\ \calo_2+m_f\ \calo_3\right]\,,
\\
&\calo_2=\frac13 {\tr}\left(\, |\phi_1|^2 + |\phi_2|^2 - 2\,|\phi_3|^2
\,\right)\,,\\
&\calo_3= -{\tr}\left( i\,\psi_1\psi_2 -\sqrt{2}g_\mt{YM}\,\phi_3
[\phi_1,\phi_1^\dagger] +\sqrt{2}g_\mt{YM}\,\phi_3
[\phi_2^\dagger,\phi_2] + {\rm h.c.}\right)\\
&\qquad\qquad +\frac23 m_f\, {\tr}\left(\, |\phi_1|^2 + |\phi_2|^2 +
|\phi_3|^2\, \right)\,.
\end{split}
\eqlabel{massdef}
\end{equation}
The non-normalizable coefficients of the gravitational bulk scalars are
the bosonic $m_b^2$ and the fermionic $m_f$ mass parameters.
The thermodynamically unstable phase of the $\caln=2^*$
plasma in $\reals^3$ is present as long as \cite{Buchel:2008uu}
\begin{equation}
0\ \leq\ \frac{m_f^2}{m_b^2}\ <\ 1\,.
\eqlabel{condition}
\end{equation}

The effective action \eqref{pw5d}
can be consistently truncated to a single scalar
$\alpha$, \ie setting
\begin{equation}
\chi\equiv 0\qquad \Longrightarrow\qquad m_f=0\,,
\eqlabel{constr}
\end{equation}
which we will do in the rest of this paper. 
To avoid cluttering we denote
\begin{equation}
m_b\equiv m\,.
\eqlabel{mbm}
\end{equation}

\subsection{PW black holes and their thermodynamics}

We use the background geometry \eqref{5dmetric} parameterization as in
\cite{Buchel:2021yay} and set
\begin{equation}
c_1=\frac{f^{1/2}}{r h^{1/4}}\,,\qquad c_2=\frac{1}{r h^{1/4}}\,,\qquad
c_3=\frac{h^{1/4}}{r f^{1/2}}\,,
\eqlabel{warps}
\end{equation}
where the radial coordinate
\begin{equation}
r\in (0,+\infty)\,.
\eqlabel{rrange}
\end{equation}
From \eqref{bac1}-\eqref{bacc}
we obtain the second order equations
\begin{equation}
\begin{split}
&0=f''-\frac{3f'}{r} -\frac{5h'f'}{4h}+4 h K\,,
\end{split}
\eqlabel{eq1}
\end{equation}
\begin{equation}
\begin{split}
&0=h''-\frac{5(h')^2}{4h}-16 h (\alpha')^2\,,
\end{split}
\eqlabel{eq2}
\end{equation}
\begin{equation}
\begin{split}
&0=\alpha''+a' \biggl(
\frac{f'}{f}-\frac3r-\frac{5h'}{4h}
\biggr)+\frac{h^{1/2}}{6r^2 f} \left(e^{2 \alpha}-e^{-4 \alpha}\right)\,,
\end{split}
\eqlabel{eq3}
\end{equation}
and the first order constraint
\begin{equation}
\begin{split}
&0=(\alpha')^2+\frac{h K}{2f}-\frac{(h')^2}{16h^2}
+\frac{h' f'}{16f h}-\frac{h'}{2r h}+\frac{f'}{4r f}-\frac{1}{r^2}
+\frac{h^{1/2}}{6r^2 f} \left(e^{2 \alpha}+\frac12 e^{-4 \alpha}\right)\,.
\end{split}
\eqlabel{eqc}
\end{equation}
Eqs.~\eqref{eq1}-\eqref{eqc} are solved with the following asymptotics:
\nxt in the UV, \ie as $r\to 0$
\begin{equation}
\begin{split}
&f=1+16 r^2 K-32 r^3 \beta K+r^4 f_{4,0}+\calo(r^5)\,,
\end{split}
\eqlabel{uv1}
\end{equation}
\begin{equation}
\begin{split}
&h=16-64 \beta r+160 r^2 \beta^2-320 r^3 \beta^3+r^4 \biggl(
-\frac{128}{9} a_{2,0} a_{2,1}+\frac{416}{27} a_{2,1}^2+\frac{256}{3} a_{2,0}^2
+560 \beta^4\\
&+\left(\frac{512}{3} a_{2,0} a_{2,1}-\frac{128}{9} a_{2,1}^2\right)\ \ln r
+\frac{256}{3}a_{2,1}^2\ \ln^2 r \biggr)+\calo(r^5\ln^2 r)\,,
\end{split}
\eqlabel{uv2}
\end{equation}
\begin{equation}
\begin{split}
&\alpha=r^2 \left(a_{2,0}+a_{2,1}\ \ln r\right)
-\beta \left(2 a_{2,0}+a_{2,1}+2 a_{2,1}\ \ln r\right) r^3+r^4 \biggl(
3 a_{2,0} \beta^2+\frac52 a_{2,1} \beta^2\\&
-8 K a_{2,1}+a_{2,0}^2-2 a_{2,0} a_{2,1}+\frac32 a_{2,1}^2+\left(
3 a_{2,1} \beta^2+2 a_{2,0} a_{2,1}-2 a_{2,1}^2\right)\ \ln r+ a_{2,1}^2\ \ln^2 r
\biggr)\\&+\calo(r^5\ln^2 r)\,;
\end{split}
\eqlabel{uv3}
\end{equation}
\nxt in the IR, \ie as $y\equiv \frac 1r\to 0$
\begin{equation}
\begin{split}
&f=f^h_1 y-\biggl(\frac53 r_0^2 (h^h_0)^{1/2}+7 h^h_0 K+
\frac{5(h^h_0)^{1/2}}{6r_0^4}\biggr)\ y^2+\calo(y^3)\,,\\
&\hat{h}=h^h_0-\biggl(
\frac{8(h^h_0)^{3/2} r_0^2}{3f^h_1}+\frac{8 (h^h_0)^2 K}{f^h_1}
+\frac{4(h^h_0)^{3/2}}{3f^h_1 r_0^4}
\biggr)\ y+\calo(y^2)\,,\\
&\alpha=\ln r_0-\frac{(r_0^6-1) (h^h_0)^{1/2}}{6f^h_1 r_0^4}\ y+\calo(y^2)\,,
\end{split}
\eqlabel{ir}
\end{equation}
where we defined
\begin{equation}
\hat{h}\equiv y^{-4}\ h \,.
\eqlabel{defhh}
\end{equation}

The non-normalizable coefficient $a_{2,1}$ of the bulk scalar
$\alpha$ is related to the mass parameter
\eqref{mbm} as follows \cite{Buchel:2000cn,Buchel:2007vy}
\begin{equation}
m^2=\frac{3}{8} a_{2,1}\,.
\eqlabel{ma21}
\end{equation}
$\beta$ is a residual gauge parameter of the background geometry
\eqref{warps} parameterization associated with the constant $\lambda$
rescaling of radial coordinate \eqref{rrange}, $r\to r\lambda$ or
$y\to \frac{y}{\lambda}$:
\begin{equation}
\begin{split}
&\beta\to \beta\lambda\,,\qquad K\to K\lambda^2\,,\qquad a_{2,1}\to
a_{2,1}\lambda^2\,,\qquad a_{2,0}\to (a_{2,0}+a_{2,1}\ln\lambda)\lambda^2\,,\\
&f_{4,0}\to f_{4,0}\lambda^2\,,\qquad f_1^h\to \frac{f_1^h}{\lambda}\,,\qquad
h_0^h\to \frac{h_0^h}{\lambda^4}\,,\qquad r_0\to r_0\,.
\end{split}
\eqlabel{scaling2}
\end{equation}
Because \eqref{scaling2} acts on all dimensionfull parameters ($K$ and $m$
specifically), the results of the holographic renormalization of the
model has to be expressed as dimensionless quantities \eqref{defhat}. 
The holographic renormalization of the $\caln=2^*$ model has been
discussed extensively in the past, so we present the results
only \cite{Buchel:2004hw,Buchel:2012gw,Buchel:2015lla}:
\begin{equation}
\begin{split}
&\hat{\cale}=\frac{1}{24\pi^4}\biggl(-\frac23
-\frac{2f_{4,0}}{3a_{2,1}^2}+\frac{128K^2}{3a_{2,1}^2}
-\frac83 \ln 2-\frac43 \ln a_{2,1}+\frac{8a_{2,0}}{3a_{2,1}}
+\frac{32 \beta^2 K}{a_{2,1}^2}\biggr)\,,\\
&\hat{s}=\frac{2^{7/2}}{3^{5/2}\pi^3(h_0^h)^{3/4}a_{2,1}^{3/2}}\,,\qquad
\frac{T}{m}=\frac{f_1^h}{6^{1/2}\pi (h_0^h)^{1/2} a_{2,1}^{1/2}}\,.
\end{split}
\eqlabel{thermores}
\end{equation}
Note that \eqref{thermores} are left invariant under \eqref{scaling2}.
The basic thermodynamic relation,
\begin{equation}
\calf=\cale-s T\,,
\eqlabel{fes}
\end{equation}
is automatically enforced by the holographic renormalization
\cite{Buchel:2004hw}, while the first law of thermodynamics,
\begin{equation}
d\hat\cale=\frac{T}{m}\ d\hat s\ \bigg|_{\frac{K}{m^2}={\rm const}}\,,
\eqlabel{firstlaw}
\end{equation}
must be verified numerically. We always check \eqref{firstlaw}
in numerical constructions of the $\caln=2^*$ black brane/black hole
geometries --- a sample of tests, for $\frac{K}{m^2}=0$ and
$\frac{K}{m^2}=1$, is shown in fig.~\ref{fl}.

\subsection{Helicity $h=0$ QNMs of the PW black holes}\label{pwbhgen}

We suppress the $h=0$ superscript, and refer to a single scalar index
as $s=0$, rather then $s=(0,j)$. Using the background parameterization
\eqref{warps}, we obtain from \eqref{eqms}:
\begin{equation}
\begin{split}
0=&\cald_2 F_2\  -W_{2,2}\ F_2-W_{2,0}\ F_0\,,\\
0=&\cald_2 F_0\  -W_{2,0}\ F_2-W_{0,0}\ F_0\,,\\
\end{split}
\eqlabel{qnm2}
\end{equation}
where the second-order differential operator $\cald_2$ (coming
from $\square$ on the background geometry \eqref{5dmetric}) is
\begin{equation}
\begin{split}
\cald_2 F(t,r)\equiv  -\frac{h^{1/2} r^2}{f}\ \del^2_{tt} F+\frac{r^2 f}{h^{1/2}}\
\del^2_{rr}F+\biggl(
\frac{r^2f'}{h^{1/2}}-\frac{5r^2 f h'}{4h^{3/2}}-\frac{3 r f}{h^{1/2}}
\biggr)\ \del_r F-{h^{1/2} r^2 k^2 }\ F\,,
\end{split}
\eqlabel{defcald}
\end{equation}
and
\begin{equation}
\begin{split}
&W_{2,2}=-\frac{1024 r^4 h^{7/2} f k^2 (3 K-k^2) (\alpha')^2}
{G^2}
+\frac{32 r^2 h^{3/2} f (h' r+4 h)^2 (k^2-3 K)^2}
{G^2}\\
&-\frac{2r (3 K-k^2)}{3G h^{1/2}} \biggl(
16 h^3 k2 r^2+6 h h' f' r^2
-9 f (h')^2 r^2+24 h^2 f' r-72 h f h' r-144 h^2 f
\biggr)\\
&-\frac43 k^2 h^{1/2} r^2\,,
\end{split}
\eqlabel{w22pw}
\end{equation}
\begin{equation}
\begin{split}
&W_{2,0}=\frac{8\sqrt{2k^2 (k^2-3K)}}{3G^2}   \biggl(
9 h^{1/2} f r^2 (h' r+4 h) (16 K h^2 r-8 h^2 k^2 r
+4 h f'+h' f' r) \alpha'\\
&+2 h^2 r (8 h^2 k^2 r-12 h f'-3 h' f' r) \left(e^{2 \alpha}-e^{-4 \alpha}\right)
-576 h^{5/2} f f' r^4 (\alpha')^3
\biggr)\,,
\end{split}
\eqlabel{w20pw}
\end{equation}
\begin{equation}
\begin{split}
&W_{0,0}=-\frac{1}{3h G^2} \biggl(
96 h^2 f' r \left(e^{2 \alpha}-e^{-4 \alpha}\right) (8 h^2 k^2 r-12 h f'
-3 h' f' r) \alpha'+288 h^{1/2} r^2 (\alpha')^2\\
&\times \biggl(32 K h^4 f k^2 r^2
-16 h^4 f k^4 r^2+8 h^3 (f')^2 k^2 r^2-24 h^3 f f' k^2 r
-6 h^2 f h' f' k^2 r^2\\
&-12 h^2 (f')^3 r
-3 h h' (f')^3 r^2+48 h^2 f (f')^2
+24 h f h' (f')^2 r+3 f (h')^2 (f')^2 r^2\biggr) 
\\&-13824 h^{5/2} f (f')^2 r^4 (\alpha')^4\biggr)
-\frac{1}{3}\left( e^{2 \alpha}+2 e^{-4 \alpha}\right)\,,
\end{split}
\eqlabel{w00pw}
\end{equation}
with $k$ given by \eqref{ks3} and
\begin{equation}
G\equiv 8 h^2 k^2 r-12 h f'-3 h' f' r\,.
\eqlabel{defG}
\end{equation}

Generically, $F_0$ and $F_2$, as well as $\ww=\ww(\kk)$, are complex.
We need to impose the normalizable boundary conditions as $r\to 0$,
and the incoming wave boundary conditions at the black brane/black hole
horizon, \ie as $y\equiv \frac 1r\to 0$. We can explicitly factor
the boundary conditions, and the harmonic time dependence,
redefining $F_0$ and $F_2$ as
\begin{equation}
\begin{split}
&F_0(t,r)=(1+r)^{i \ww/2}\ \frac{r^2}{1+r^2}\ e^{-i 2\pi T\ww t}\ f_0(r)\,,\\
&F_2(t,r)=(1+r)^{i \ww/2}\ \frac{r^2}{1+r^2}\ e^{-i 2\pi T\ww t}\ f_2(r)\,,
\end{split}
\eqlabel{redeffs}
\end{equation}
which renders both $f_0(r)$ and $f_2(r)$
regular and $\calo(1)$ both at the boundary and at the horizon.
While the boundary normalization of the master scalar $F_2\sim \calo(r^2)$
near the AdS$ _5$ boundary is independent of any
presence of the bulk gravitational scalars, the boundary
normalization of the master scalar $F_0\sim \calo(r^2)$
reflects the fact that this scalar is associated with the
fluctuations of the gravitational bulk scalar $\alpha$ dual to an operator
of the conformal dimension $\Delta=2$. The incoming wave boundary
condition at the horizon, \ie $F\propto y^{-i\ww/2}$ is universal for all
the master scalars. Further introducing
\begin{equation}
f_0=f_{0,\Re}+ i f_{0,\Im}\,,\qquad f_2=f_{2,\Re}+ i f_{2,\Im}\,,\qquad
\ww=\ww_{\Re}+i \ww_{\Im}\,,
\eqlabel{splitrei}
\end{equation}
we obtain from \eqref{qnm2} four (generically coupled)
second order linear ODEs for
\begin{equation}
\{f_{0,\Re}\,,\  f_{0,\Im}\,,\, f_{2,\Re}\,, f_{2,\Im}\}\,.
\eqlabel{setfs}
\end{equation}
Because of the linearity, there is an arbitrary complex overall
normalization of solution; we fix this normalization imposing
\begin{equation}
\begin{split}
&\ell\ge 2:\qquad \lim_{r\to 0} f_{2,\Re}=1\,,\qquad \lim_{r\to 0} f_{2,\Im}=0\,,\\
&\ell<2:\qquad \lim_{r\to 0} f_{0,\Re}=1\,,\qquad \lim_{r\to 0} f_{0,\Im}=0\,.
\end{split}
\eqlabel{normalization}
\end{equation}
The QNM equations for \eqref{setfs} are solve with the following asymptotics:
\nxt in the UV, \ie as $r\to 0$ (without loss of generality,
we work in the gauge $\beta=1$ -- see \eqref{uv1}-\eqref{uv3})
\begin{equation}
\begin{split}
&f_{2,\Re}=1+\left(-2+\frac12 \ww_\Im\right)\ r+\calo(r^2)\,,\qquad
f_{2,\Im}=-\frac12 \ww_\Re\ r+\calo(r^2)\,,\\
&f_{0,\Re}=f_{0,\Re,0}+\left(-2 f_{0,\Re,0}+\frac12 f_{0,\Im,0}\ \ww_\Re
+\frac12 f_{0,\Re,0}\ \ww_\Im\right)\ r+\calo(r^2)\,,\\
&f_{0,\Im}=f_{0,\Im,0}+\left(-2 f_{0,\Im,0}+\frac12
f_{0,\Im,0}\ \ww_\Im-\frac12 f_{0,\Re,0}\ \ww_\Re\right)\ r+\calo(r^2)\,;
\end{split}
\eqlabel{qnmuv}
\end{equation}
\nxt in the IR, \ie as $y\equiv \frac 1r\to 0$
\begin{equation}
\begin{split}
&f_{2,\Re}=f_{2,\Re,0}^h+\calo(y)\,,\qquad
f_{2,\Im}=f_{2,\Im,0}^h+\calo(y)\,,\\
&f_{0,\Re}=f_{0,\Re,0}^h+\calo(y)\,,\qquad
f_{0,\Im}=f_{0,\Im,0}^h+\calo(y)\,.
\end{split}
\eqlabel{qnmir}
\end{equation}
Note that, for a fixed background and $\kk$,
the solution is characterized in total by 8 parameters
\begin{equation}
\left\{\ \ww_\Re\,,\ \ww_\Im\,,\ f_{0,\Re,0}\,,\ f_{0,\Im,0}\,,\
f_{2,\Re,0}^h\,,\ f_{2,\Im,0}^h\,,\ f_{0,\Re,0}^h\,,\ f_{0,\Im,0}^h
\
\right\}\,,
\eqlabel{parqnm}
\end{equation}
precisely as needed to specify a solution of 4 second order ODEs. 

We now highlight the reduction of the set of the QNM equations in
some special cases.
\begin{itemize}
\item For branches of the QNM modes with $\Re[\ww]=0$, as in
figs.~\ref{hydroK0}, \ref{l2}, and \ref{higherl} we can 
consistently set 
\begin{equation}
f_{2,\Im}(r)\equiv 0\,,\qquad f_{0,\Im}(r)\equiv 0\,,\qquad \ww_\Re=0 \,,
\eqlabel{red1}
\end{equation}
correspondingly eliminating, in addition to $\ww_\Re$, 
$f_{0,\Im,0}$, $f_{2,\Im,0}^h$, and $f_{0,\Im,0}^h$ from the parameter list \eqref{parqnm} -- we are left with 2 second order differential equations and
4 specifying parameters.
\item Special case harmonics $\ell=0$ and $\ell=1$ can be treated by
dropping the master equation for $f_2$, and setting $k=0$ or $k=\sqrt{3 K}$
(for $\ell=0$ and $\ell=1$ correspondingly) in the remaining equations
for $f_{0,\Re}$ and $f_{0,\Im}$. Note that in this case we need to normalize
the solution as in the second line in \eqref{normalization}. QNM branches
of this type appear in figs.~\ref{f0mode1K0}, \ref{f0mode2K0} and
\ref{f0mode2K0}.
\item Yet further reduction occurs for $\ell=0$ and $\ell=1$ if the
QNMs have $\Re[\ww]=0$, as in figs.~\ref{f0mode1K0} and \ref{l0and1}.
Here we have a single second order equation for the function
$f_{0,\Re}$ (with $k=0$ or $k=\sqrt{3 K}$) and a pair of specifying
parameters $\ww_\Im$ and $f_{0,\Re,0}^h$.
\item We included $K=0$ (the black brane) case of QNMs at $k=0$ with
$\ell=0$ case covered above. 
\end{itemize}

We add  some practical remarks for the QNM
computation.
\begin{itemize}
\item In solving the QNM boundary value problems we use the shooting method
developed in \cite{Aharony:2007vg}.
\item To obtain results for the QNMs reported in figs.~\ref{f0mode1K0}
and \ref{f0mode2K0} we set $K=0$, and produce the data sets of backgrounds
changing in small increments
parameter $a_{2,1}$ in \eqref{uv1}-\eqref{uv2} from $a_{2,1}=0$ (the pure AdS$ _5$-Schwarzschild black brane) to the value $a_{2,1}=0.09158879$.
We fix the gauge parameter $\beta=1$. This allows a black brane
construction covering the range of $\frac{m^2}{T^2}$ reported in
figs.~\ref{macro} and \ref{csbv}. Note that a monotonic change in $a_{2,1}$
translates into non-monotonic dependence of $\frac{m^2}{T^2}$, allowing to
cover both the thermodynamically stable and the unstable phases.
The thermodynamically unstable state of interest \eqref{tm} is within the
scanned range. For each black brane background we compute the QNMs, using
as initial seeds the AdS$ _5$-Schwarzschild black brane
QNMs reported in fig.~\ref{adsqnm}.
\item  To obtain results for the QNMs reported in figs.~\ref{l0and1}-\ref{l2}
we produce $\caln=2^*$ black hole backgrounds keeping
mass-to-temperature ratio $\frac{m^2}{T^2}$ fixed at \eqref{tm}, and changing
$K$ from zero in small increments.
For each generated black hole
background we compute the QNMs, using
as initial seeds the QNMs of the $\caln=2^*$ black brane. 
Since we work in
a fixed $\beta=1$ gauge, the physical curvature is not $K$, but rather its
dimensionless analog, 
$\frac{K}{m^2}$. Similar to the black brane constructions from variation
of $a_{2,1}$, we observe that while we monotonically increase $K$
at fixed $\frac{m^2}{T^2}$, $\frac{K}{m^2}$ varies non-monotonically ---
this is the origin of the non-single valuedness of the QNM spectra in
figs.~\ref{l0topbot} and \ref{notsingle}.
\end{itemize}

\subsection{Helicity $h=0$ QNMs of PW black branes in the limit
$a_{2,1}\to 0$}

The purpose of this section is to discuss the limit of the
general QNM equations \eqref{qnm2} when $K=0$ and $a_{2,1}=0$.
This is a pure AdS$ _5$-Schwarzschild black brane limit,
and could be helpful to the reader to study prior to tackling
the general case covered in section \ref{pwbhgen}. 

Solving \eqref{eq1}-\eqref{eqc} with $K=0$, $a_{2,1}=0$
and the boundary conditions \eqref{uv1}-\eqref{ir} we find:
\begin{equation}
\begin{split}
\alpha\equiv 0\,,\qquad h=\frac{16}{(\beta r+1)^4}\,,\qquad
f=\frac{(2 \beta r+1) (2 \beta^2 r^2+2 \beta r+1)}{(\beta r+1)^4}\,.
\end{split}
\eqlabel{adseqs}
\end{equation}
As in section \ref{pwbhgen}, we set the gauge parameter
$\beta=1$. 

Given \eqref{adseqs}, the general QNM equations \eqref{qnm2} decouple:
\begin{equation}
\begin{split}
0=&\del^2_{tt} F_0-\frac{(2 r+1)^2 (2 r^2+2 r+1)^2}{16(r+1)^4}\
\del^2_{rr} F_0\\&
-\frac{(2 r+1) (2 r^2+2 r+1) (4 r^4-10 r^2-10 r-3)}{16r (r+1)^5}\
\del_r  F_0\\
&+\frac{(2 r+1) (2 r^2+2 r+1) (4 k^2 r^2-r^2-2 r-1)}
{4(r+1)^4 r^2}\ F_0\,,
\end{split}
\eqlabel{qnmads0}
\end{equation}
\begin{equation}
\begin{split}
0=&\del^2_{tt} F_2-\frac{(2 r+1)^2 (2 r^2+2 r+1)^2}{16(r+1)^4}\
\del^2_{rr} F_2
\\&-\frac{(2 r+1) (2 r^2+2 r+1) (4 r^4-10 r^2-10 r-3)}{16r (r+1)^5}\
\del_r F_2\\
&+\frac{k^2  (2 r+1) (2 r^2+2 r+1)}{(r+1)^4
(8 k^2 r^2+16 k^2 r+3 r^2+8 k^2)^2 r^2}
\biggl(
64 k^4 r^6+256 k^4 r^5-16 k^2 r^6+384 k^4 r^4\\&-96 k^2 r^5-15 r^6
+256 k^4 r^3-240 k^2 r^4-72 r^5+64 k^4 r^2-320 k^2 r^3
-108 r^4\\&-240 k^2 r^2-72 r^3-96 k^2 r-18 r^2-16 k^2\biggr)\ F_2\,.
\end{split}
\eqlabel{qnmads2}
\end{equation}

\begin{figure}[t]
\begin{center}
\psfrag{k}[cc][][1][0]{$\kk$}
\psfrag{i}[tt][][1][0]{$\Im[\ww]$}
\psfrag{r}[bb][][1][0]{$|\Re[\ww]|$}
\includegraphics[width=3in]{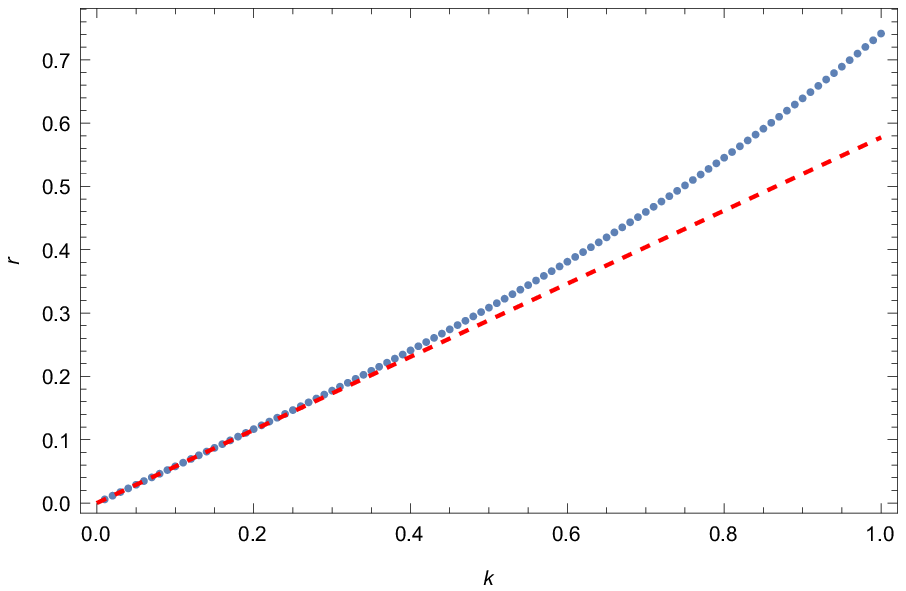}
\includegraphics[width=3in]{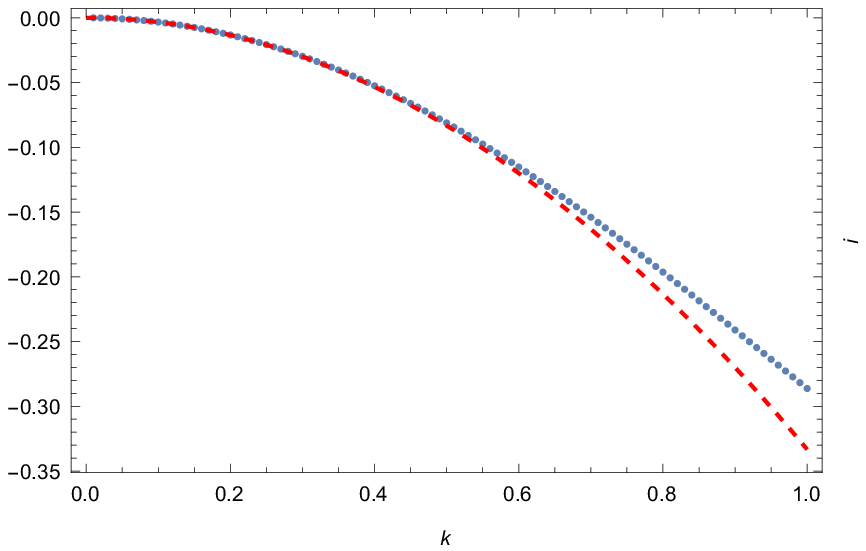}
\end{center}
  \caption{The hydrodynamic QNM of AdS$ _5$-Schwarzschild black brane
  computed in the framework of appendix \ref{eomshol}. Dashed red curves
represent the leading $\kk\to 0$ approximation, see \eqref{cftlim}.} \label{cftri}
\end{figure}

\begin{itemize}
\item Eq.~\ref{qnmads0} has a smooth limit $k\to 0$. Redefining $F_0(t,r)$
as in \eqref{redeffs} and solving the resulting equations
at $k=0$, we produce the spectrum of $\Delta=2$ QNMs in AdS$ _5$-Schwarzschild
black brane reported in fig.~\ref{adsqnm}, reproducing the appropriate
results of \cite{Nunez:2003eq}.
\item Because of the factor $(8 k^2 r^2+16 k^2 r+3 r^2+8 k^2)$ in the
denominator of the equation \eqref{qnmads2}, the $k\to 0$ and $r\to 0$
limits do not commute. We can still solve, say for the hydrodynamic
QNM, using the redefinition \eqref{redeffs} and generate the spectrum
$\ww(\kk)$ for $\kk\ge \frac{1}{100}$. The results are presented in
fig.~\ref{cftri}, along with the expected small $\kk$
holographic-CFT approximation (dashed red curves):
\begin{equation}
\ww(\kk)=\pm \frac{1}{\sqrt{3}}\ \kk-\frac 13\ \kk^2+\calo(\kk^3)\,.
\eqlabel{cftlim}
\end{equation}
At $\kk=1$ we can compare our results with those reported in appendix E of
\cite{Jansen:2019wag} (JRR):
\begin{equation}
\begin{split}
{\rm this\ work}:\qquad &\ww(1)=\pm 0.7414299655(2)-0.2862800072(6)\ i\,,\\
{\rm JRR}:\qquad &\ww(1)=\pm 0.7414299655(?)-0.2862800072(?)\ i\,,
\end{split}
\eqlabel{compare}
\end{equation}
where $(?)$ denotes the unreported truncation in \cite{Jansen:2019wag}.
\end{itemize}

\section{Conclusions}\label{conclude}

In this paper we extended the master field framework of
black brane/black hole  QNM computations to Einstein-multi-scalar
gravitational effective actions of the type \eqref{efac}. 

As an application, we discussed the stabilization of the
thermodynamically unstable $\caln=2^*$ black brane horizons,
as the boundary gauge theory is compactified on $S^3$ of a curvature
scale $K$. The initial instability in the helicity $h=0$ hydrodynamic sector, present
for  small curvature,  is cured once $K> K_{\ell=2}^{unstable}$. Surprisingly,
we found that while $\ell=0$  QNMs are stable  for
$K\in [0,K_{\ell=2}^{unstable}]$, they develop an instability for yet larger
$K>K_{\ell=0}^{stable}>K_{\ell=2}^{unstable}$. They can ultimately be stabilized for
$K>K_{\ell=0}^{unstable}$. $\ell=1$ QNMs in the helicity $h=0$ sector
in the model discussed are always stable.

In the example of the $\caln=2^*$ holography, the extended thermodynamically
unstable horizons can be alternatively stabilized when the space of the
boundary gauge theory is compactified on sufficiently small three-torus:
the unstable hydrodynamic modes are simply projected from
the spectrum\footnote{We would like to thank David Mateos for pointing
this out.}. However, toroidal compactification would not be able to
cure the instability of the holographic
conformal order, present in the non-hydrodynamic sector at zero
spatial momentum \cite{Buchel:2020jfs}. 

Many avenues are left open for future studies:
\nxt it would be interesting to understand when Klebanov-Strassler
black holes \cite{Buchel:2021yay} become dynamically stable;
\nxt it would be interesting to explore whether the holographic conformal
order \cite{Buchel:2020thm,Buchel:2020xdk,Buchel:2020jfs}
can be stabilized on $S^d$, before it is destroyed;
\nxt what aspects of the stabilization when black branes
are {\it} deformed to black holes are universal? are $\ell=1$
harmonics in the helicity $h=0$ sector always stable?

Finally, in this work we considered effective actions
relevant to holographic models without the bulk
gauge field, correspondingly without conserved global charges of the thermal
states of the boundary gauge theory.
It is only in this setting that one straightforwardly relate via \eqref{cscv}
the thermodynamic and the dynamical instabilities \cite{Buchel:2005nt}.
In the presence of a conserved $U(1)$ charge
density $\rho$ and a chemical potential $\mu$   the story is more
nuanced\footnote{See \cite{Buchel:2010wk}.},
as the expression for the speed of the sound waves \eqref{cscv} becomes
more complicated in the charged plasma: 
\begin{equation}
c_s^2=\biggl(
(\cale+P)\ \frac{\del(P,\rho)}{\del(T,\mu)}+\rho\
\frac{\del(\cale,P)}{\del(T,\mu)}
\biggr)\biggl((\cale+P)\ \frac{\del(\cale,\rho)}{\del(T,\mu)}\biggr)^{-1}\,.
\eqlabel{cscharged}
\end{equation}
Additionally, in the presence of the global charges, one can no longer
establish in full generality the stability of the QNMs in the helicity
$h=1$ sector. It would be interesting to explore this in the future.

\section*{Acknowledgments}
This research is supported in part by Perimeter Institute for
Theoretical Physics.  Research at Perimeter Institute is supported in
part by the Government of Canada through the Department of Innovation,
Science and Economic Development Canada and by the Province of Ontario
through the Ministry of Colleges and Universities. This work was
further supported by NSERC through the Discovery Grants program.

\appendix
\section{Master equations of Einstein-scalar black holes in $D=5$}\label{eomshol}
In this section we extend the work of \cite{Jansen:2019wag} to black holes
in theories of Einstein gravity in $D=5$ space-time dimensions with multiple
scalars\footnote{The motivation is to verify and fix typos in the original work  \cite{Jansen:2019wag},
and to prepare the stage for the stability analysis of the
black holes on the conifold with fluxes \cite{Buchel:2021yay}.}.
We adhere to the notations of \cite{Jansen:2019wag} as much as it is practical. 

Consider an effective action
\begin{equation}
S_5=\int_{\calm_5}d^{3+2}\xi\ \sqrt{-g} \biggl[R-\sum_{j=1}^p \eta_j \left(\del\phi_j\right)^2-V\left(\{\phi_j\}\right)\biggr]\,,
\eqlabel{efac}
\end{equation}
where $j=1\cdots p$ indices the scalars $\phi_i$; $\eta_i$'s are the constant normalizations of the scalar
kinetic terms, and $V$ is an arbitrary potential. We will be interested in the stability analysis of the
black branes/holes in the theory \eqref{efac} with maximally symmetric 3-dimensional Schwarzschild horizons:
\begin{equation}
ds_5^2=-c_1^2\ dt^2+ c_2^2\ dX_{3,K}^2+c_3^2\ dr^2\,,
\eqlabel{5dmetric}
\end{equation}
where $c_i=c_i(r)$, $\phi_j=\phi_j(r)$ and 
\begin{equation}
dX_{3,K}^2=\begin{cases}
d\bm{x}^2\equiv dx_1^2+dx_2^2+dx_3^2\,,\qquad &K=0\,,\qquad {\rm planar}\,,\\
d\Omega_{(3)}^2\,,\qquad  &K>0\,,\qquad {\rm spherical}\,,\\
dH_{(3)}^2\,,\qquad  &K<0\,,\qquad {\rm hyperbolic}\,.
\end{cases}
\eqlabel{defxnk}
\end{equation}
Note that we do not fix $K=\{0,\pm 1\}$, but instead allow it to vary smoothly --- this would allow
for the interpolation of the quasinormal spectra between different maximally symmetric horizons, notably
between the planar and the spherical ones.
A useful way to explicitly parameterize $X_{3,K}$ and a metric on it is as follows:
$X_{3,K}=(x_1\equiv x, x_2\equiv y,  x_3\equiv z)$, and 
\begin{equation}
dX_{3,K}^2=\frac{dx^2}{(1-K x^2)}+(1-K x^2)\ \biggl[\frac{dy^2}{(1-K y^2)}+(1-K y^2)\ dz^2\biggr]\,.
\eqlabel{dx3k}
\end{equation}

From \eqref{efac} we obtain the following second order equations of motion ($'\equiv \frac{d}{dr}$  and
$\del_j\equiv \frac{\del}{\del\phi_j}$)
\begin{equation}
0=c_1''+c_1' \left[\ln\frac{c_2^3}{c_3}\right]'+\frac{c_3^2 c_1}{3}\ V\,,
\eqlabel{bac1}
\end{equation}
\begin{equation}
0=c_2''+c_2' \left[\ln\frac{c_1 c_2^2}{c_3}\right]'+\frac{c_2 c_3^2}{3}\ \left(V-\frac{6 K}{c_2^2}\right)\,,
\eqlabel{bac2}
\end{equation}
\begin{equation}
0=\phi_j''+\phi_j' \left[\ln\frac{c_1 c_2^3}{c_3}\right]'-\frac{c_3^2}{2\eta_j}\
\del_jV\,,
\eqlabel{bac3}
\end{equation}
and the first order constraint\footnote{We verified that \eqref{bacc} is consistent
with \eqref{bac1}-\eqref{bac3}.}
\begin{equation}
0=\sum_{j=1}^p\eta_j (\phi_j')^2-\left[\ln c_2^3\right]' \left[\ln(c_1^2 c_2^2)\right]'
+c_3^2\ \left(\frac{6 K}{c_2^2}-V\right)\,.
\eqlabel{bacc}
\end{equation}

We organize all the gauge invariant fluctuations into three sets of master scalars of different helicity $h$:
\begin{itemize}
\item the helicity $h=2$ set, $\{\Phi_2^{(2)}\}$;
\item the helicity $h=1$ set, $\{\Phi_2^{(1)}\}$;
\item the helicity $h=0$ set, $\{\Phi_2^{(0)},\Phi_{(0,j)}^{(0)}\}$, $j=1\cdots p$.
\end{itemize}
Any master scalar $\Phi_s^{(h)}$ ($s=2$  or $s=(0,j)$ and $h=\{0,1,2\}$) is assumed to
have the following dependence:
\begin{equation}
\Phi_s^{(h)}(\xi)=F_{s}^{(h)}(t,r)\ S(X_{3,K})\,,
\eqlabel{masterscalar}
\end{equation}
where $S(X_{3,K})$ is a scalar eigenfunction of the Laplacian $\Delta_K$ on \eqref{dx3k} with an eigenvalue
$k^2$:
\begin{equation}
\Delta_K\ S + k^2\ S=0 \,.
\eqlabel{compact}
\end{equation}
In this work we will be concerned with planar ($K=0$) or  spherical  ($K>0$)  horizons. In the former
case, $k^2 \in [0,+\infty)$ and in the latter case
\begin{equation}
k^2=K \ell (\ell +2)\qquad   {\rm with}\qquad   \ell\in \zet_+\,.
\eqlabel{ks3}
\end{equation}

Each of the master scalars satisfies a coupled master equation of the form
\begin{equation}
\square \Phi_s^{(h)}-W_{s,s'}^{(h)}(r)\ \Phi_{s'}^{(h)}=0\,,
\eqlabel{eqms}
\end{equation}
where $\square$ is the wave operator on the full $D=5$ metric \eqref{5dmetric}, and the symmetric potential
matrix,
\begin{equation}
W_{s,s'}^{(h)}=W_{s',s}^{(h)}\,,
\eqlabel{sym}
\end{equation}
couples master scalars in a given helicity set.

We now present results for potentials $W_{s,s'}^{(h)}$ in different helicity sectors,
as well as relations between the  master scalars and a specific set of gauge invariant
fluctuations in that sector. We refer the reader to \cite{Jansen:2019wag} for a detailed
discussion exactly how the gauge invariant fluctuations are
constructed\footnote{In a holographic setting gauge invariant fluctuations in Einstein-scalar theories
were used for the first time in \cite{Benincasa:2005iv}.}. 

\subsection{Helicity $h=0$ sector}

The relation between the gauge invariant fluctuations in the scalar sector
\begin{equation}
\{\f0_j\,,\ \h2_{tr}\,,\ \h2_{rr}\,,\ \h2_{xr}\,,\ \h2_{tt}\}\,,
\eqlabel{gauge0}
\end{equation}
and the master scalars
\begin{equation}
\{F^{(0)}_{(0,j)}\,,\ F^{(0)}_2\}\,,
\eqlabel{gauge0m}
\end{equation}
all being functions of $(t,r)$, is as follows:
\begin{equation}
\begin{split}
\f0_j&=-\frac{1}{\sqrt{2\eta_j}}\ F_{(0,j)}^{(0)}+\frac{\sqrt3 k\ c_2\phi_j'}
{6\kt\ c_2'} \ F_2^{(0)} \,,
\end{split}
\eqlabel{h01}
\end{equation}
\begin{equation}
\begin{split}
&\h2_{rr}= \frac{ c_2 c_3^2  (c_2' c_1-c_1' c_2)}{D}\ \sum_{j=1}^p\bigg\{\sqrt{2\eta_i}\phi_j'\
F_{(0,j)}^{(0)}\bigg\}
+\biggl(
\frac{c_2^2 c_3^2\ k}{3\sqrt{3}(c_2')^2\ \kt}\ \sum_{j=1}^p \left\{\eta_j (\phi_j')^2\right\}
\\&-\frac{c_2 c_3^2  c_1'\ k}{\sqrt{3} c_1 c_2'\ \kt }
-\frac{2\sqrt{3} c_1 c_3^4 K\  k }{D\ \kt}-\frac{k\ \sqrt{3}c_3^2\left((c_2')^2 c_1-c_3^2c_1k^2
-c_1' c_2' c_2\right)}{\kt\ D}
\biggr)\ F_2^{(0)}\\
&+\frac{c_2 c_3^2\  k }{\sqrt{3} c_2'\ \kt}\ \del_r F_2^{(0)}\,,
\end{split}
\eqlabel{h02}
\end{equation}
\begin{equation}
\begin{split}
\h2_{xr}&=\frac{c_2^2c_3^2c_1 }{D}\ \sum_{j=1}^p \left\{ \sqrt{\frac{\eta_j}{2}}\phi_j'\  F_{(0,j)}^{(0)}\right\}
+\biggl(
-\frac{3\sqrt{3} c_1 c_2 c_2' c_3^2  K}{k \kt\ D}
+ \frac{c_2 c_3^2\ k   }{2\sqrt{3}c_2'D\ \kt }\biggl(c_3^2  c_1k^2\\&
+3 (c_2')^2 c_1+3 c_1' c_2' c_2\biggr)\biggr)\ F_2^{(0)}
+\frac{\sqrt{3} c_2^2}{2k \kt}\  \del_rF_2^{(0)}\,,
\end{split}
\eqlabel{h03}
\end{equation}
where we set
\begin{equation}
\kt\equiv \sqrt{k^2-3K}\,,\qquad D\equiv
c_3^2  c_1k^2-3 (c_2')^2 c_1+3 c_1' c_2' c_2\,.
\eqlabel{h0set}
\end{equation}
Additionally,
\begin{equation}
\begin{split}
&\h2_{tt}=\frac{2  c_1^2}{c_3^2} \left(\ln\frac{c_3}{c_1c_2}\right)'\ \h2_{xr}
-\frac{2 c_1^2}{c_3^2}\ \del_r\h2_{xr}+\frac{c_1^2}{c_3^2}\ \h2_{rr}\,,\\
&\h2_{tr}=\frac{4 c_2^2}{k^2}\ \sum_{j=1}^p\bigg\{
\eta_j\phi_j'\ \del_t \f0_j \bigg\}
+2 \del_t\h2_{xr}-\frac{6 c_2' c_2 }{c_3^2\ k^2}\ \del_t\h2_{rr}\,.
\end{split}
\eqlabel{h04}
\end{equation}

The master scalars $\Phi^{(0)}_s$ satisfy \eqref{eqms} with the potentials:
\begin{equation}
\begin{split}
&W_{2,2}^{(0)}=\frac{4}{3D^2 c_2^2} \biggl(
c_2^2 c_1^2 c_3^2\ k^2 \tk^2\ \sum_{n=1}^p \left\{\eta_n (\phi_n')^2\right\}
+(k^2-6 K) D^2-9\tk^2\  c_1' c_2'c_2  D \\
&-3\tk^2(k^2-2 K)\ c_1 c_3^2 D+2\tk^4\ c_1 c_3^2
\left(c_3^2 k^2 c_1+3 c_1' c_2' c_2\right)
\biggr)\,,
\end{split}
\eqlabel{w22}
\end{equation}
\begin{equation}
\begin{split}
&W_{2,(0,i)}^{(0)}=-\frac{\tk  k\  c_3^2 c_1 }{D}\
\sqrt{\frac{2}{3\eta_i}}\ \del_iV
-\frac{\tk k\ c_2 c_1  (c_1' c_2-c_2' c_1)}{D^2}\  \sqrt{\frac{8\eta_i}{3}}\ \phi_i'\ 
\sum_{n=1}^p \left\{\eta_n (\phi_n')^2\right\}
\\&
+\frac{\tk k\  c_1 c_2'  (D-c_3^2 k^2 c_1) }{c_2 D^2}\  \sqrt{\frac{8\eta_i}{3}}\ \phi_i'
+\frac{\tk k\ c_1^2     c_3^2 c_2'(k^2-2 K)}{D^2c_2}\  \sqrt{24\eta_i}\ \phi_i'\,,
\end{split}
\eqlabel{w201}
\end{equation}
\begin{equation}
\begin{split}
&W_{(0,i),(0,j)}^{(0)}=\frac{1}{2 \sqrt{\eta_i} \sqrt{\eta_j}}\ \del_i\del_jV
-\frac{c_2 (c_2' c_1-c_1' c_2)}{D} \left(
\sqrt{\frac{\eta_j}{\eta_i}}\ \phi_j'\ \del_iV
+\sqrt{\frac{\eta_i}{\eta_j}}\ \phi_i'\ \del_jV\right)
\\&+\frac{2 c_2^2 (c_2' c_1-c_1' c_2)^2}{c_3^2 D^2}\
\sqrt{\eta_i}\phi_i'\ \sqrt{\eta_j}\phi_j'\  \sum_{n=1}^p \left\{\eta_n (\phi_n')^2\right\}
+\frac{\sqrt{\eta_i}\phi_i'\ \sqrt{\eta_j}\phi_j'}{D^2}
\biggl(2 c_1^2 c_3^2\ k^2 (k^2-2 K)\\
&+2k^2\ (c_2' c_1+2 c_1' c_2) (c_2' c_1-c_1' c_2)
-\frac{12 c_2' (c_2' c_1+c_1' c_2)
(c_2' c_1-c_1' c_2)^2}{c_3^2 c_1}\biggr)\,,
\end{split}
\eqlabel{w0102}
\end{equation}
where $\tk$ and $D$ are defined in \eqref{h0set}.

Notice that for $K>0$, the relations  \eqref{h02} and \eqref{h03}  between the master scalars
and the gauge invariant fluctuations are singular for $\ell=0$ and $\ell=1$.  
This is so because in these cases the dynamical degrees of freedom are those of the
scalars only. These cases must be treated separately \cite{Jansen:2019wag}.

\subsubsection{$\ell=0$}

For $\ell=0$, there are no $h_{tx}$, $h_{xr}$ and $h_-$ components of the metric perturbations;
furthermore, the gauge transformations can be used to set metric components $h_{tr}=0$ and $h_+=0$.
Thus, we are left with perturbations
\begin{equation}
\{\f0_j\,,\ h_{tt}\,,\ h_{rr}\}\,.
\eqlabel{h0l0}
\end{equation}
We find\footnote{As pointed out in \cite{Jansen:2019wag}, here, as well as for $\ell=1$,
there is also a certain inhomogeneous
piece in the master equations. This piece must be set to zero to study  fluctuations
in a fixed-mass black hole background.}
\begin{equation}
\begin{split}
&h_{rr}=\frac{2c_3^2  c_2}{3c_2'}\ \sum_{i=1}^p \left\{\eta_i \phi_i' \f0_i\right\}\,,
\end{split}
\eqlabel{l01}
\end{equation}
\begin{equation}
\begin{split}
c_1^2\ \del_r\left(\frac{h_{tt}}{c_1^2}\right)=&
-\frac{2 c_1^2 c_2 }{3c_2'}\   \sum_{i=1}^p\left\{\eta_i\phi_i'\ \del_r\f0_i\right\}
+\frac{c_1^2 c_3^2 c_2 }{3c_2'} \sum_{i=1}^p\left\{\del_iV\ \f0_i\right\}
\\&+\frac{2 c_1^2 c_2^2} {9(c_2')^2}\
\sum_{j=1}^p\left\{\eta_j(\phi_j')^2\right\}\ \sum_{i=1}^p\left\{\eta_i \phi_i'\  \f0_i\right\}
-\frac{4 c_1 (c_1 c_2'+c_2 c_1')}{3c_2'}
\ \sum_{i=1}^p\left\{ \eta_i\phi_i' \ \f0_i\right\}\,.
\end{split}
\eqlabel{l02}
\end{equation}
Introducing the master scalars as 
\begin{equation}
\f0_j=-\frac{1}{\sqrt{2\eta_j}}\ F^{(0)}_{(0,j)}\,,
\eqlabel{h0l0def}
\end{equation}
we obtain master equations for $\Phi^{(0)}_{(0,j)}$ with $W^{(0)}_{(0,i),(0,j)}$
given formally by \eqref{w0102} in the limit $k\to 0$.

\subsubsection{$\ell=1$}

For $\ell=1$ there is no $h_-$ component of the metric fluctuations. 
We can use gauge transformations to set metric components $h_{tx}=0$ and
$h_+=0$. Thus, are left with the (gauge variant) perturbations 
\begin{equation}
\{\f0_j\,,\ h_{tr}\,,\ h_{rr}\,,\ h_{xr}\,,\ h_{tt}\}\,.
\eqlabel{gauge01}
\end{equation}
Note that the scalar eigenfunction for $\ell=1$ is explicitly
\begin{equation}
S(X_{3,K})=x \sqrt{K}\,.
\eqlabel{sl1}
\end{equation}
From the fluctuation equations of motion we find (compare with \eqref{h04})
\begin{equation}
\begin{split}
&h_{tt}=\frac{2  c_1^2}{c_3^2} \left(\ln\frac{c_3}{c_1c_2}\right)'\ h_{xr}
-\frac{2 c_1^2}{c_3^2}\ \del_rh_{xr}+\frac{c_1^2}{c_3^2}\ h_{rr}\,,\\
&h_{tr}=\frac{4 c_2^2}{k^2}\ \sum_{j=1}^p\bigg\{
\eta_j\phi_j'\ \del_t \f0_j \bigg\}
+2 \del_t h_{xr}-\frac{6 c_2' c_2 }{c_3^2\ k^2}\ \del_t h_{rr}\,.
\end{split}
\eqlabel{htr}
\end{equation}

The remaining fluctuations can be expressed through the master scalars \eqref{gauge0m} as
\begin{equation}
\begin{split}
\f0_j&=-\frac{1}{\sqrt{2\eta_j}}\ F_{(0,j)}^{(0)}+\frac{g\ c_2\phi_j'}
{c_2'} \ {F}_2^{(0)} \,,
\end{split}
\eqlabel{h01l1}
\end{equation}
\begin{equation}
\begin{split}
&h_{rr}= \frac{ c_2 c_3^2  (c_2' c_1-c_1' c_2)}{D}\ \sum_{j=1}^p\bigg\{\sqrt{2\eta_i}\phi_j'\
F_{(0,j)}^{(0)}\bigg\}
+2g \biggl(
\frac{c_2^2 c_3^2}{3(c_2')^2}\ \sum_{j=1}^p \left\{\eta_j (\phi_j')^2\right\}
\\&-\frac{c_2 c_3^2  c_1'}{c_1 c_2'}
-\frac{6 c_1 c_3^4 K }{D}-\frac{3c_3^2\left((c_2')^2 c_1-c_3^2c_1k^2
-c_1' c_2' c_2\right)}{D}
\biggr)\ F_2^{(0)}+2g\ \frac{c_2 c_3^2 }{ c_2'}\ \del_r F_2^{(0)}\,,
\end{split}
\eqlabel{h02l1}
\end{equation}
\begin{equation}
\begin{split}
h_{xr}&=\frac{c_2^2c_3^2c_1 }{D}\ \sum_{j=1}^p \left\{ \sqrt{\frac{\eta_j}{2}}\phi_j'\  F_{(0,j)}^{(0)}\right\}
+2g \biggl(
-\frac{9 c_1 c_2 c_2' c_3^2  K}{k^2\ D}
+ \frac{c_2 c_3^2   }{2c_2'D }\biggl(c_3^2  c_1k^2\\&
+3 (c_2')^2 c_1+3 c_1' c_2' c_2\biggr)\biggr)\ F_2^{(0)}
+g\ \frac{{3} c_2^2}{k^2}\  \del_rF_2^{(0)}\,,
\end{split}
\eqlabel{h03l1}
\end{equation}
where $D$ is given by \eqref{h0set}, and $g$ is an arbitrary constant parameter of the unfixed
gauge transformation  \cite{Jansen:2019wag}.
Note that \eqref{h01l1},  \eqref{h02l1} and \eqref{h03l1} are equivalent to  \eqref{h01l1}, \eqref{h02} and \eqref{h03} up to
replacement
\begin{equation}
2g\ \longleftrightarrow\ \frac{k}{\sqrt3\ \tk}\,.
\eqlabel{replacementh0}
\end{equation}

The only physical master equations are those for the scalars $\Phi^{(0)}_{(0,j)}$
(in these equations there is a decoupling of the 'gauge' master scalar $\Phi^{(0)}_2$), with the
relevant potentials $W^{(0)}_{(0,i),(0,j)}$ obtained setting $\tk=0$ in \eqref{w0102}.
The equation for the 'gauge' master scalar can also be obtained from the general expressions
valid for $\ell\ge 2$, provided we identify (compare \eqref{h01} and  \eqref{h01l1})
\begin{equation}
F_2^{(0)}\bigg|_{\ell\ge 2}\equiv 2g\ \frac{\sqrt3\ \tk}{k}\ F_2^{(0)}\bigg|_{\ell=1}\,,
\eqlabel{replacementf2}
\end{equation}
prior to taking the limit $\tk \to 0$.
Because of \eqref{replacementf2}, this latter equation is necessarily singular in the limit $g\to 0$.

\subsection{Helicity $h=1$ sector}

The relation between the gauge invariant fluctuations in the vector sector
\begin{equation}
\{\h2_{tz}\,,\ \h2_{zr}\}\,,
\eqlabel{gauge1}
\end{equation}
and the master scalar
\begin{equation}
\{F^{(1)}_2\}\,,
\eqlabel{gauge1m}
\end{equation}
all being functions of $(t,r)$, is as follows:
\begin{equation}
\begin{split}
&\h2_{tz}=\frac{3 c_2 c_1 c_2'}{\tk\ c_3}\ F^{(1)}_2+\frac{c_2^2 c_1}{\tk\ c_3}\ \del_r F^{(1)}_2\,,
\\
&\h2_{zr}=\frac{c_3 c_2^2}{\tk\ c_1}\ \del_tF^{(1)}_2 \,,
\end{split}
\eqlabel{h12}
\end{equation}
where we deliberately introduced a singularity at $\ell=1$\, \ie $\kt=0$ \eqref{h0set},  
to highlight the fact that the fluctuations \eqref{gauge1} are gauge invariant only for $\ell\ge 1$
\cite{Jansen:2019wag}. 

The master scalar $\Phi^{(1)}_2$ satisfies \eqref{eqms} with the potential
\begin{equation}
W^{(1)}_{2,2}=\frac{1}{c_3^2}\ \sum_{j=1}^p \left\{\eta_j(\phi_j')^2\right\}-
\frac{3 K}{c_2^2}+\frac{3 (c_2')^2}{c_2^2 c_3^2}-\frac{6 c_1'c_2'}
{c_1 c_2 c_3^2}\,.
\eqlabel{w122}
\end{equation}

\subsubsection{$\ell=1$}

The case $\ell=1$ is special since in this case there are no dynamical degrees of freedom
\cite{Jansen:2019wag}. 
Indeed, here, there are no $h_{xz}$ component of the metric fluctuations,  and the gauge can be fixed
setting $h_{zr}=0$.
The remaining metric component $h_{tz}$ satisfies
\begin{equation}
\del_r \left(\frac{h_{tz}(t,r)}{c_2^2}\right)=0\,,
\eqlabel{htz}
\end{equation}
which is normalizable only if it identically vanishes.

\subsection{Helicity $h=2$ sector}

The relation between a particular  gauge invariant fluctuation $\h2_{yz}(t,r)$ and the
master  scalar $F_2^{(2)}(t,r)$ in the tensor sector is as follows 
\begin{equation}
\delta ds_5^2=\delta g_{yz}(t,r,X_{3,K})\ dydz ={\h2_{yz}(t,r)}\ \frac{S_T(x)}{1-K y^2}\ dydz=
c_2^2 F_2^{(2)}(t,r)\ \frac{S_T(x)}{1-K y^2}\ dydz  \,,
\end{equation}
where $S_T(x)$ satisfies \cite{Jansen:2019wag}
\begin{equation}
(1-K x^2)\ S_T''=-K x\ S_T'-\left(k^2+2 K \left(1+\frac{K x^2}{1-K x^2}\right)\right)\ S_T\,.
\eqlabel{sneq}
\end{equation}

The master scalar
\begin{equation}
\Phi_2^{(2)}(\xi)\equiv F_2^{(2)}(t,r)\ S(X_{3,K}) 
\eqlabel{msh2}
\end{equation}
equation is {\it universally} that of the minimally coupled massless scalar
\begin{equation}
\square \Phi_2^{(2)}=0\,,
\eqlabel{eqms2}
\end{equation}
\ie
\begin{equation}
W^{(2)}_{2,2}\equiv 0\,.
\eqlabel{w222}
\end{equation}
The universality of the $h=2$ sector at $K=0$ was emphasized originally in \cite{Buchel:2004qq}.

\subsection{Stability of $h=1$ and $h=2$ sectors}

It was argued in \cite{Jansen:2019wag} that the QNMs of Einstein-scalar
black holes/black branes, \ie when $K\ge 0$,  are stable in
$h=2$ and $h=1$ helicity sectors.

Since $h=2$ sector is not directly sensitive to scalars of the effective
action, it is obviously stable even for the more general effective 
actions \eqref{efac}. The potential of the QNMs
in the $h=1$ helicity sector explicitly depends on the scalars, see
\eqref{w122}. Nonetheless, the arguments of  \cite{Jansen:2019wag}
can still be literally repeated, leading to the conclusion of the
stability.

\section{Numerical tests}\label{err}

\begin{figure}[t]
\begin{center}
\psfrag{a}[cc][][0.7][0]{$a_{2,1}$}
\psfrag{e}[cc][][0.7][0]{$T ds/d\cale-1$}
\includegraphics[width=3in]{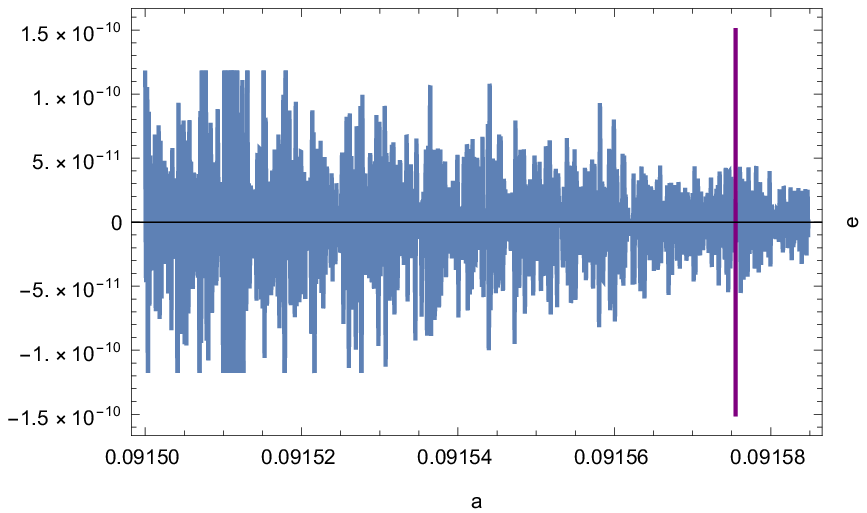}
\includegraphics[width=3in]{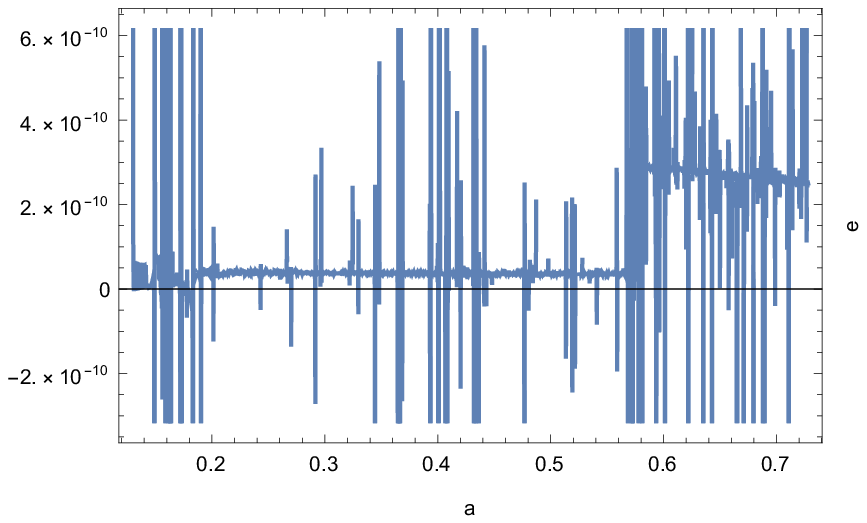}
\end{center}
  \caption{Verification of the first law of thermodynamics \eqref{firstlaw}
  for the $\caln=2^*$ black brane, $\frac{K}{m^2}=0$ (the left panel),
  and for a sample $\caln=2^*$ black hole, $\frac{K}{m^2}=1$
  (the right panel). The vertical purple line identifies
  the thermodynamically unstable state of interest \eqref{tm}, further
  used in the QNM analysis.
} \label{fl}
\end{figure}

The work presented in this paper is numerical. It is imperative that
we do as many tests as possible to confirm the reliability of the
results. In the rest of this section we highlight a  subset of
the tests that we performed.

\begin{itemize}
\item The  $\caln=2^*$ black brane thermodynamics has been discussed
previously in \cite{Buchel:2007vy}. The previous work is done in a
different parameterization of the background geometry, compare to the
parameterization \eqref{warps} used here. We
recover\footnote{Our current numerical algorithms are more precise.}
the previous results.
\item The hydrodynamics of $\caln=2^*$ black brane was discussed earlier
(again in a different parameterization) 
in \cite{Buchel:2007mf,Buchel:2008uu}. Once again, we recover
the previous results.
\item The AdS$ _5$-Schwarzschild $\Delta=2$ black brane spectrum at $\kk=0$
presented in fig.~\ref{adsqnm} agrees with the results reported
in \cite{Nunez:2003eq}, obtained using a completely different method.
\item While we did not present the AdS$ _5$-Schwarzschild results
for the $h=0$ sector graviton QNMs\footnote{See however \cite{talk}.},
obtained solving for the QNM spectrum from \eqref{qnmads2} at
$\kk=\frac{1}{100}$, we confirmed that the results obtained are in
agreement with $\Delta=4$ results reported
in \cite{Nunez:2003eq}, obtained using a completely different method.
\item We use the master field formalism \cite{Jansen:2019wag},
but we use finite difference rather than the spectral methods in computing
the QNMs, as well as a different background geometry parameterization.
We find agreement with the numerical result for the
AdS$ _5$-Schwarzschild  $h=0$ graviton QNM at $\kk=1$, see  \eqref{compare}.
\item In constructing $\caln=2^*$ black brane and black hole geometries, we
always verify the first law of thermodynamics \eqref{firstlaw}.
Typical results of such tests are shown in fig.~\ref{fl}.
\item $\caln=2^*$ black hole background at temperature
\eqref{tm} and a given value of $\frac{K}{m^2}$ can be reached from
AdS$ _5$-Schwarzschild horizon geometry in two ways:
\nxt we can keep $K=0$ and increase $\frac{m^2}{T^2}$ from zero to
\eqref{tm}, followed by the increase of $\frac{K}{m^2}$ from zero to the
value of interest;
\nxt we can start with AdS$ _5$-Schwarzschild black hole at a corresponding
value of $\frac{K}{T_p^2}$, and increase $\frac{m^2}{T^2}$ from zero
to \eqref{tm}\\
--- in both ways we land at an identical geometry.
\item The hydrodynamic QNM of the $\caln=2^*$ black brane in the
thermodynamically unstable state \eqref{tm} reported in fig.~\ref{hydroK0}
is in excellent agreement with the prediction \eqref{sound},
using the appropriate equilibrium transport coefficients \eqref{transport}.
\end{itemize}

\bibliographystyle{JHEP}
\bibliography{n2curved}

\end{document}